\documentclass[twocolumn, tighten]{aastex631}
\usepackage{bm}
\usepackage{amsmath} %数学公式
\usepackage{booktabs}
\usepackage{array}
\usepackage{enumerate}
\usepackage{url}
\newcommand{\re}[1]{#1}
\newcommand{\rep}[1]{#1}

\newcommand{\KIAA}{\affiliation{Kavli Institute for Astronomy and
Astrophysics, Peking University, Beijing 100871, China}}
\newcommand{\DOA}{\affiliation{Department of Astronomy, School of Physics,
Peking University, Beijing 100871, China}}

\newcommand{\NAOC}{\affiliation{National Astronomical Observatories,
Chinese Academy of Sciences, Beijing 100012, China}}

\received{XXX}
\revised{YYY}
\accepted{ZZZ}
\submitjournal{ApJ}
\shorttitle{Multiband BNS/NSBH}
\shortauthors{C. Liu and L. Shao}
\begin{document}

\title{Neutron Star--Neutron Star and Neutron Star--Black Hole Mergers: \\
Multiband Observations and Early Warnings}
\correspondingauthor{Lijing Shao}
\email{lshao@pku.edu.cn}
\author[0000-0001-7649-6792]{Chang Liu}\DOA\KIAA
\author[0000-0002-1334-8853]{Lijing Shao}\KIAA\NAOC

%---------------------------------------------------------------------

\begin{abstract}
The detections of gravitational waves (GWs) from binary neutron star (BNS)
systems  and  neutron star--black hole (NSBH) systems provide new insights into
dense matter properties in extreme conditions and associated high-energy
astrophysical processes.  However, currently information about NS equation of
state (EoS) is extracted with very limited precision.  Meanwhile, the fruitful
results from the serendipitous discovery of the $\gamma$-ray burst alongside
GW170817 show the necessity of early warning alerts.  Accurate measurements of
the matter effects and sky location could be achieved by joint GW detection from
space and ground.  In our work, based on two example cases, GW170817 and
GW200105, we use the Fisher information matrix analysis to investigate the
multiband synergy between the space-borne decihertz GW detectors and the
ground-based Einstein Telescope (ET). We specially focus on the parameters
pertaining to spin-induced quadrupole moment, tidal deformability, and sky
localization.  We demonstrate that, (i) only with the help of multiband
observations can we constrain the quadrupole parameter; and (ii) with the
inclusion of decihertz GW detectors, the errors of tidal deformability would be
a few times smaller, indicating that many more EoSs could be excluded; (iii)
with the inclusion of ET, the sky localization improves by about an order of
magnitude. Furthermore, we have systematically compared the different limits
from four planned decihertz detectors and adopting two widely used waveform
models. 
\end{abstract}

%---------------------------------------------------------------------

\keywords{Gravitational wave astronomy (675) 
--- Neutron stars (1108) 
--- Gravitational wave detectors (676) 
--- Compact binary stars (283)}

%---------------------------------------------------------------------

\allowdisplaybreaks  % allow long equations to display on different pages

%---------------------------------------------------------------------
\section{Introduction}
\label{sec:intro}
%---------------------------------------------------------------------

Until now, more than 50 gravitational wave (GW) events have been published by
the LIGO/Virgo Collaboration \citep{LIGOScientific:2018mvr,
LIGOScientific:2020ibl, LIGOScientific:2021usb}, in which the majority is from
binary black hole (BBH) mergers. In comparison, the GW signals from binary
neutron star (BNS) systems \citep{LIGOScientific:2017vwq,LIGOScientific:2020aai}
and neutron star--black hole (NSBH) systems \citep{LIGOScientific:2021qlt} are
rare but of special interests, as they could help us comprehend high-density
nuclear matter \citep{LIGOScientific:2018cki}, improve views about astrophysical
processes under extreme conditions \citep{LIGOScientific:2017pwl}, and
understand compact object populations \citep{LIGOScientific:2020kqk}. 

Extracting BNS and NSBH properties solely from GW signals is crucial for GW
astronomy, which highly depends on the accuracy of the waveform. Two dominant
finite-size effects distinguish NSs from BHs: (i) the deformation due to NS's
own rotation, and (ii) due to the companion's tidal field. They enter the
waveform as self-spin term \citep{Poisson:1997ha} and tidal term
\citep{Flanagan:2007ix,Vines:2011ud} respectively. With the accurate waveform
model \citep{Dietrich:2019kaq,Dietrich:2018uni}, we could constrain the
equation-of-state (EoS) dependent spin-induced quadrupole moment and tidal
deformability, and pick out the correct EoS model \citep{Read:2009yp,
Hinderer:2009ca, Agathos:2015uaa}, thus informing the low-energy quantum
chromodynamics and quark confinement behaviours.  Moreover, we could test the
nature of BHs \citep{Krishnendu:2019tjp, Narikawa:2021pak}, distinguish BNS
models from BBH models \citep{Chen:2020fzm, Gralla:2017djj, Krishnendu:2018nqa},
and test alternative gravity theories~\citep{Shao:2017gwu, Sennett:2017lcx,
Shao:2019gjj}.

In addition to the GW signal, short $\gamma$-ray burst (GRB), GRB\,170817A was
found right after the peak of the first BNS inspiral, GW170817
\citep{LIGOScientific:2017vwq, LIGOScientific:2017zic}. Together with the
following counterparts in X-ray, ultraviolet, optical, infrared, and radio
bands, simultaneous detections of GWs and electromagnetic (EM) signals initiate
a new era of multi-messenger astronomy with precious information
\citep{LIGOScientific:2017ync}. In the meantime, EM signals also call for a
better localization ability from GW detectors. Scientists have explored the
future localization abilities of LIGO/Virgo detectors
\citep{Nitz:2020vym,Magee:2021xdx}, as well as the third generation (3G)
detectors including the Europe-led Einstein Telescope \citep[ET;][]{Hild:2010id}
and the US-led Cosmic Explorer \citep[CE;][]{Evans:2016mbw}, using the
post-Newtonian (PN) waveform \citep{Zhao:2017cbb,Chan:2018csa} with precession
\citep{Tsutsui:2020bem}, eccentricity \citep{Ma:2017bux,Pan:2019anf}, and tidal
effects \citep{Wang:2020xwn}.

For the discovered LIGO/Virgo sources, the angular resolution of the 3G GW
detectors can be as accurate as a few degrees~\citep[see e.g.,
][]{Zhao:2017cbb}.  On the other hand, the space-borne detectors could localize
sources within arcminutes \citep{Takahashi:2003wm,Nair:2018bxj}. To obtain even
better constraints, the multiband observations could be a win-win solution for 
both space-borne and ground-based detectors. 

The BNS and NSBH signals can hardly reach the signal-to-noise ratio (SNR)
threshold of the millihertz-band space-borne detectors such as LISA
\citep{Audley:2017drz}, and will spend more than a few years before coalescence
even if they do. Therefore we direct our attention on the decihertz detectors,
e.g., Decihertz Observatories \citep[DOs;][]{Sedda:2019uro, ArcaSedda:2021dte}
and DECihertz laser Interferometer Gravitational wave Observatory
\citep[DECIGO;][]{Yagi:2011wg,Kawamura:2018esd}.  Because of their shorter arm
length, decihertz detectors are sensitive in the frequency range of
0.01--10\,Hz. DOs have two LISA-like proposals, the ambitious DO-Optimal and the
less challenging DO-Conservative. DECIGO also has two designs. B-DECIGO is a
primordial version of DECIGO consisting of one LISA-like detector, while the
complete design of DECIGO consists of four independent LISA-like detectors and
uses Fabry-Perot cavity to achieve a much lower noise level.

As shown in early studies, the joint detection of BNSs and NSBHs with decihertz
detectors and ground-based detectors will improve the parameter precision
prominently \citep{Nakamura:2016hna, Nair:2015bga, Liu:2020nwz, Nakano:2021bbw}.
\citet{Isoyama:2018rjb} and \citet{Nair:2018bxj} have shown the precision
improvement specially focusing on BNS finite-size effects and the angular
resolution. In this work, we extend the study in \citet{Isoyama:2018rjb} by
constraining both finite-size effects and localization parameters
simultaneously, for both BNS and NSBH systems. 
\re{Comparing to previous works, we use the updated sensitivity curves and detector designs. 
For the first time, we give the parameter errors and its multiband improvement distributions on the sky maps.
Due to the need of early warnings, we further include multiband sky localization as a function of time. Moreover, our work gives a systematical analysis on parameter correlations, illustrates the capability of different detectors, and compares the implementation of different waveforms.
}
Our work enables a better understanding of joint observations, and
could provide more information for different observing scenarios. 

In this work, with the help of the Fisher matrix analysis, we investigate the
multiband measurement uncertainties considering the complete parameter space
including spin, tidal, self spin and location parameters. We give the sky
distributions of multiband enhancement for quadrupole-monopole parameters, tidal
deformabilities and angular resolutions, as well as the pre-merger localization
precision as a function of inspiraling time. We compare the parameter estimation
(PE) results of BNS and NSBH systems, using the ET, as a representative of 3G
ground-based GW detectors, jointly with a decihertz detector, either from
B-DECIGO, DECIGO, DO-Conservative, or DO-Optimal. We adopt both PN and
phenomenological waveforms in our study. To be more instructive, we show the
projected multiband constraints on NS's EoS using the limits from tidal
deformability.  We hope such a detailed study, augmenting the existing
investigations, can help researchers lay out the near-future detector science
objectives more clearly and understand better about the depth of NS physics
that we will learn from such kinds of multiband observations.

The organization of the paper is as follows.  In Sec.~\ref{sec:method} we
introduce the method used in our work, where Sec.~\ref{sec:wf} reviews the NS
waveform models; Sec.~\ref{sec:det} provides the detectors' configurations and
responses; and Sec.~\ref{sec:pe method} briefly summarizes the Fisher matrix
method and the source properties under study.  In Sec.~\ref{sec:result} we
present our complete PE results, where Sec.~\ref{sec:corr} displays the
parameter correlations for BNS and NSBH systems; Sec.~\ref{sec:int} shows the
multiband improvement of quadrupole-monopole parameter, tidal deformability, as
well as the differences between BNS and NSBH systems; Sec.~\ref{sec:ext} focuses
on limits of extrinsic parameters, especially on the sky localization precision
and early warning alerts; Sec.~\ref{sec:other_det} compares the PE measurements
given by different decihertz detectors; and Sec.~\ref{sec:PV2} compares the
limits imposed by using PN waveform and phenomenological waveform. In
Sec.~\ref{sec:EoS} we discuss constraints on the NS's EoS, and finally in
Sec.~\ref{sec:sum} we briefly summarize our work.  Throughout this paper we use
geometrized units in which $G=c=1$.

%---------------------------------------------------------------------
\section{Method}
\label{sec:method}
%---------------------------------------------------------------------

In this section, we first introduce the waveforms used in the following
calculations in Sec.~\ref{sec:wf}, then we introduce the detectors we use and
their responses to GWs in Sec.~\ref{sec:det}, and at last in Sec.~\ref{sec:pe
method}, we briefly summarize the PE method and list the physical properties of
the specific example systems that we explore.

%---------------------------------------------------------------------
\subsection{Waveform Construction}
\label{sec:wf}
%---------------------------------------------------------------------

We model the GW signal using the Fourier domain, restricted PN approximation
\citep{Buonanno:2009zt}. With Fourier representation computed using the
stationary phase approximation, the source-frame strain is
%--
\begin{align}
\tilde{h}_{+}(f)&=\mathcal{A} f^{-7 / 6} e^{i \Psi_{\mathrm{NS}}(f)}\,, \\
\tilde{h}_{\times}(f)&=\mathcal{A} f^{-7 / 6} e^{i[ \Psi_{\mathrm{NS}}(f)+\frac{\pi}{2}]}\,, 
\label{eq:source_wf}
\end{align}
%--
where the amplitude $\mathcal{A}= \sqrt{{5}/{24}} \pi^{-2 / 3} {\mathcal{M}^{5 /
6}}/{D_{L}}$, in which $D_{{L}}$ is the luminosity distance of the source and
${\cal M}=M\eta^{3/5}$ is the chirp mass with the total mass $M=m_1 + m_2$ and
the symmetric mass ratio $\eta=m_1m_2/M^2$.  Due to the cosmological expansion,
we measure the redshifted mass {$m_{1,2} =(1+z)\,m_{1,2}^{\rm S}$} of the two
compact objects, where $z$ is the redshift calculated from $D_{{L}}$, and
$m_{1,2}^{\rm S}$ are the source-frame component masses with $m_{1}^{\rm S} \geq
m_{2}^{\rm S}$ by default.  Note that we include amplitude's dependence on the
inclination angle in the pattern function (see Sec.~\ref{sec:det}).  

The phase $\Psi_{\mathrm{NS}}(f)$ in the waveform is,
%--
\begin{align}
 \Psi_{\mathrm{NS}}(f) &= 2 \pi f t_{\mathrm{c}}-2\phi_{\mathrm{c}}-\frac{\pi}{4}  \nonumber \\ 
 & \hspace{-0.3cm} +\frac{3}{128 \eta v^{5}}\big(\Psi_{3.5 \,
 \mathrm{PN}}^{\mathrm{pp}+\mathrm{spin}}+ \Psi_{2\mbox{--}3.5
 \,\mathrm{PN}}^{\mathrm{QM}}+ \Psi_{5\mbox{--}6 \, \mathrm{PN}}^{\mathrm{ tidal
 }}\big)\,,
\label{eq:phase}
\end{align}
%--
where $t_c$ and $\phi_c$ are the time and orbital phase at coalescence, and $v =
(\pi M f)^{1/3}$ is the orbital velocity. Note that terms with ${\cal O}
\big(v^{2n}\big)$ correspond to the $n_{\rm th}$ PN order.

Apart from the BBH baseline waveform, we consider two matter effects specially
generated by NSs: the spin-induced and tidal-induced deformations, which
respectively count for the second and third terms in the bracket of
Eq.~(\ref{eq:phase}). In total, the GW phase contains three parts, as elaborated
below. 
%--
\begin{enumerate}[(i)]
  \item The point particle term, $\Psi_{3.5 \,
  \mathrm{PN}}^{\mathrm{pp}+\mathrm{spin}}$ \citep{Arun:2008kb,Mishra:2016whh},
  is kept up to 3.5\,PN.  Because we consider the non-precessing case,
  $\Psi_{3.5 \, \mathrm{PN}}^{\mathrm{pp}+\mathrm{spin}}$ also contains the
  aligned spin effect characterized by the dimensionless spin parameters
  projected to the angular momentum direction, $\chi_{i} = \boldsymbol{S}_i
  \cdot \hat{\boldsymbol{L}}/(m_i^{\rm S})^2$, where $\boldsymbol{S}_i$ is
  the spin angular momentum and $\hat{\boldsymbol{ L}}$ is the unit normal of
  the orbital plane, expressed later in Eq.~\eqref{eq:angular}.  $\Psi_{3.5 \,
  \mathrm{PN}}^{\mathrm{pp}+\mathrm{spin}}$ includes the linear spin-orbit
  effects up to 3.5\,PN order, quadratic-in-spin (spin-spin) effects to 3\,PN
  order, and cubic-in-spin (spin-spin-spin) effects to the (leading) 3.5\,PN
  order. 
  %---
  \item The second term is the quadrupole-monopole term, $\Psi_{2\mbox{--}3.5 \,
  \mathrm{PN}}^{\mathrm{QM}}$.    The spin-induced quadrupole moment $Q_i =
  -\kappa_i\chi_i^2(m_i^{\rm S})^3$ is a measure of the degree of the oblateness
  due to NS's rotation, where $\kappa_i$ is a dimensionless quadrupole parameter
  with $\kappa_i \sim 2\mbox{--}20$ for NSs and $\kappa_i = 1$  for BHs
  \citep{Narikawa:2021pak}. This finite size effect that depends on NS's EoS
  enters the GW signal as an order-$v^4$ correction through the
  quadrupole-monopole interaction  \citep{Poisson:1997ha,Mikoczi:2005dn} and we
  include them up to 3.5\,PN order by \citep{Krishnendu:2017shb, Nagar:2018plt,
  Dietrich:2019kaq},
	%--
	\begin{align}  
	\Psi_{2\mbox{--}3.5 \, \mathrm{PN}}^{\mathrm{QM}} &= -25 \tilde{\mathrm{Q}}
	v^{4} \nonumber\\ 
	&+\bigg\{\Big(\frac{15635}{42}+60 \eta\Big) \tilde{\mathrm{Q}}
	-\frac{2215\delta_M}{24}  \delta \tilde{\mathrm{Q}}\bigg\} v^{6}
	\nonumber\\ 
	&+\bigg\{\Big[\frac{375}{2}\left(\chi_{s}+\chi_{a}\delta_M\right)  - 200
	\pi-10 \eta \chi_{s} \Big]
	\tilde{\mathrm{Q}} \nonumber\\ 
	& \quad\quad -\frac{1985}{6}\left(\chi_{a}+\chi_{s}\delta_M \right)
	\delta \tilde{\mathrm{Q}}\bigg\} v^{7} \,,
	\end{align}
	%--
  where
	\begin{align}
		\tilde{Q} &= \frac{ 2m_1^2\chi_1^2(\kappa_1-1) +
		2m_2^2\chi_2^2(\kappa_2-1) } {M^2}\,,  \\
		\delta\tilde{Q} &= \frac{ -2m_1^2\chi_1^2(\kappa_1-1) +
		2m_2^2\chi_2^2(\kappa_2-1) } {M^2}\,,
	\end{align}
  %--
  with $\delta_M = (m_1-m_2)/M$ and $\chi_{s,a} = (\chi_1 \pm
  \chi_2)/2$.  It is $\tilde{Q}$, the combination of individual quadrupole
  parameters $\kappa_i$, to which GW detectors are most sensitive, while
  $\delta \tilde Q$ is the subdominant parameter. We find that, to
  simultaneously constrain $\kappa_1$ and $\kappa_2$,  or $\tilde Q$ and $\delta
  \tilde Q$, is difficult due to the strong degeneracies among the quadrupole
  parameters and the spin parameters. Therefore we only constrain the leading
  term $\tilde Q$, and we will refer to it as the ``quadrupole term'' in the
  following analyses.
  %---
  \item The last term is the tidal term $\Psi_{5\mbox{--}6 \,
  \mathrm{PN}}^{\mathrm{tidal}}$ \citep{Flanagan:2007ix,Vines:2011ud}. At the
  last stages of the inspiral, the quadrupolar tidal field $\mathcal{E}_{i j}$
  of one compact object induces a quadrupole moment ${Q}_{i j}$ to the other
  component. To the leading order in the adiabatic approximation, ${Q}_{i
  j}=-\lambda\mathcal{E}_{i j}$ where $\lambda$ is the tidal Love number which
  takes the form $\lambda(m_i^{\rm S},{\rm EoS})  = 2k_2R^5(m_i^{\rm S})/3$,
  with $k_2$ being the second Love number, and $R(m_i^{\rm S})$ is the NS radius
  as a function of its mass. Both $k_2$ and $R(m_i^{\rm S})$ are EoS dependent.
  The deformation effect enters the GW phase from 5\,PN through the
  dimensionless tidal deformability parameter $\Lambda_i = \lambda/(m_i^{\rm
  S})^5$. We also include the next-to-leading order (6\,PN) term
  \citep{Wade:2014vqa,Narikawa:2021pak}, then
	%--
	\begin{equation}
		\hspace{-0.4cm} \Psi_{5\mbox{--}6 \, \mathrm{PN}}^{\mathrm{tidal}} =-\frac{39}{2} \tilde{\Lambda} v^{10}+\Big(-\frac{3115}{64} \tilde{\Lambda}+\frac{6595\delta_M}{364}  \delta \tilde{\Lambda}\Big) v^{12}\,,
	\end{equation}
	%--
	where the combined dimensionless tidal deformabilities $\tilde\Lambda$ and
	$\delta\tilde\Lambda$ are
	{\allowdisplaybreaks
	\begin{align}
		\tilde{\Lambda} & =\frac{8}{13} \bigg[ \Big(1+7 \eta-31
		\eta^{2}\Big)\left(\Lambda_{1}+\Lambda_{2}\right) \nonumber\\
		&+\Big(1+9 \eta-11
		\eta^{2}\Big)\left(\Lambda_{1}-\Lambda_{2}\right)\delta_M\bigg]\,, \\ 
		\delta \tilde{\Lambda} &=\frac{1}{2}\bigg[\Big(1-\frac{13272}{1319}
		\eta+\frac{8944}{1319}
		\eta^{2}\Big)\left(\Lambda_{1}+\Lambda_{2}\right)\delta_M \nonumber\\
		& \hspace{-0.7cm} +\Big(1-\frac{15910}{1319}
		\eta+\frac{32850}{1319} \eta^{2}+\frac{3380}{1319}
		\eta^{3}\Big)\left(\Lambda_{1}-\Lambda_{2}\right)\bigg]\,.
	\end{align}
	}
	%--
	Similar to $\tilde Q$, the tidal phase is dominated by the leading term
	characterized by $\tilde\Lambda$, and the contribution from $\delta\tilde
	\Lambda$ is small. Hence, we exclude the estimation of $\delta\tilde
	\Lambda$ in our work. We refer to $\tilde\Lambda$ as the ``tidal
	deformability'' of the system throughout this work. It is worth noting that
	BHs have zero tidal deformability \citep{Binnington:2009bb}, and for
	asymmetric NSBHs ($\Lambda_1$ = 0, $m_2 \ll m_1$) or very massive BNSs
	($\Lambda_1$, $\Lambda_2 \rightarrow 0$), $\tilde\Lambda$ will be small and
	thus they would be indistinguishable from BBHs ($\tilde\Lambda$ = 0). 
	\rep{For equal mass systems, $\delta\tilde\Lambda$=0.} 
\end{enumerate}
%--

Essentially, in matched filtering, since we do not know the true EoS, we search
for the quadrupole parameters and tidal parameters independently. Nevertheless,
with the universal Q-Love relations \citep{Yagi:2013awa,Yagi:2016bkt}, one can
prescribe the quadrupole moments through the tidal deformability without the
knowledge of the correct EoS, therefore reducing the number of parameters to
infer.  Because our purpose is to constrain the EoS, we use $\tilde Q$ and
$\tilde\Lambda$ as separate parameters to estimate. There are waveforms that use
the universal relation, such as the phenomenological waveforms
\citep{Dietrich:2018uni, Dietrich:2019kaq}, which we will discuss in
Sec.~\ref{sec:PV2}. In that specific section, we will constrain only
$\tilde\Lambda$.

%---------------------------------------------------------------------
\subsection{Detector Responses and Sensitivities}
\label{sec:det}
%---------------------------------------------------------------------

After having the source-frame waveform in the last subsection, we now construct
the detector responses and obtain the detector-frame waveform. For the
space-borne detectors, we use the method in Sec.~2.1 of \citet{Liu:2020nwz} to
model their responses.  The basic idea is as follows. The signal received by the
detector is
%--
\begin{align}
\tilde{h}(f) &= \Big[ F^{+}\big(  f \big) \, \tilde{h}_{+}(f) +F^{ \times} \big( f \big) \, \tilde{h}_{ \times}(f)\Big] e^{-i{\varphi_D}(f)}\,,
\label{eq:det_response}
\end{align}
%--
where the location dependent pattern functions are,
%--
\begin{widetext}
\begin{align}
	F^{+}(\theta, \phi, \psi,\iota)  &=\frac{\big(1+\cos ^{2} \iota\big)}{2}\ 
	\Big[\frac{1}{2} \big(1+\cos ^{2} \theta \big) \cos 2 \phi \cos 2 \psi
	 -\cos \theta \sin 2 \phi \sin 2 \psi \Big] \,, \label{eq:Fplus} \\
    F^{\times}(\theta, \phi, \psi,\iota) &=\cos \iota\ \Big[ \frac{1}{2}
    \big(1+\cos ^{2} \theta \big) \cos 2 \phi \sin 2 \psi 
    +\cos \theta \sin 2 \phi \cos 2 \psi \Big] \,.
    \label{eq:Fcross}
\end{align}
\end{widetext}
%--
The \{$\theta, \phi, \psi, \iota$\} are the time-varying source direction angles
($\theta$, $\phi$), polarization angle ($\psi$), and inclination angle ($\iota$)
in the detector frame. Since we know the orbital motion of the detector, the way
to construct the response is to use the fixed \{$\bar \theta_S,
\bar\phi_S,\bar\theta_L,\bar\phi_L$\}, which are the source direction and
angular momentum direction in the Solar system barycentric frame, and the time
$t$ to substitute \{$\theta, \phi, \psi,\iota$\} \citep[see details in Sec.~2 of
][]{Liu:2020nwz}.  The last term of Eq.~(\ref{eq:det_response}) is the Doppler
phase correction \citep{Cutler:1997ta},
%--
\begin{equation}
	\varphi_{D}(t)={2 \pi f}R \sin {\bar\theta}_{S} \cos
	\left[{\Phi_{\rm{space}}}(t)-\bar\phi_S\right]\,,
\end{equation}
%--
where $R = 1$\,AU is the orbital radius of the detector, and
${\Phi}_{\rm{space}}(t)$ is the azimuthal angle of the detector around the Sun.

The BNS and NSBH signals normally last more than one day in 3G ground-based
detectors, so they also have a time-varying $F^+, F^\times$ and $\varphi_D$. For
ground-based detectors, we follow the same logic as with space-borne detectors
to construct their responses. The difference between them is
in the transformation from \{$\theta, \phi, \psi,\iota$\} in
Eqs.~(\ref{eq:Fplus}--\ref{eq:Fcross}) to \{$\bar \theta_S,
\bar\phi_S,\bar\theta_L,\bar\phi_L, t\}$, which is determined by the detector
orbits. 

Ground-based detectors rotate with the Earth.  We define the latitude of the
detector $\delta$, the inclination of Earth's equator with respect to the
ecliptic plane $\epsilon=23^{\circ}26'$, the length of a sidereal day $T_E$.
The Earth's self rotation phase is ${\Phi}_{\rm{g}}(t)= {2\pi t}/{T_E} +
{\Phi}_0$, where ${\Phi}_0$ is the initial phase. By assuming that one arm
$\hat{\bm{x}}$ points to the south and the other arm $\hat{\bm{y}}$ points to
the east, then the unit detector frame
$\hat{\bm{x}}$-$\hat{\bm{y}}$-$\hat{\bm{z}}$ in ecliptic coordinate is 
%--
\begin{widetext}
{\allowdisplaybreaks
\begin{align}
	\hat{\boldsymbol x} &= \Big(\sin\delta\cos\Phi_{\rm{g}}(t), ~
	\sin\delta\sin\Phi_{\rm{g}}(t)\cos\epsilon-\cos\delta\sin\epsilon, ~
	-\sin\delta\sin\Phi_{\rm{g}}(t)\sin\epsilon-\cos\delta\cos\epsilon\Big)
	\,,\\
	\hat{\boldsymbol y} &= \Big(-\sin\Phi_{\rm{g}}(t), ~ \cos\Phi_{\rm{g}}(t)\cos\epsilon, ~ -\cos\Phi_{\rm{g}}(t)\sin\epsilon\Big)
	\,,\\
	\hat{\boldsymbol z} &= \Big(\cos\delta\cos\Phi_{\rm{g}}(t), ~
	\cos\delta\sin\Phi_{\rm{g}}(t)\cos\epsilon+\sin\delta\sin\epsilon, ~
	-\cos\delta\sin\Phi_{\rm{g}}(t)\sin\epsilon+\sin\delta\cos\epsilon\Big)
	\,,
\end{align}}
\end{widetext}
%--
in which $\hat{\boldsymbol{z}}$ points from the Earth center to the detector.
Note that, the arm direction could alter, with a rotation angle $\alpha_0$,
which describes the initial orientation of the detector arms. Together with the
the unit vector,  
%--
\begin{equation}
	\hat{\boldsymbol L} = \left(\sin\bar\theta_L
	\cos\bar\phi_L, ~ \sin\bar\theta_L\sin\bar\phi_L, ~
	\cos\bar\theta_L\right)\,,
	\label{eq:angular}
\end{equation}
which is the direction of orbital angular momentum of
the source, and the unit vector,
%--
\begin{equation}
	\hat{\boldsymbol N} = \left(\sin\bar\theta_S\cos\bar\phi_S, ~
	\sin\bar\theta_S\sin\bar\phi_S, ~ \cos\bar\theta_S\right)\,,
\end{equation} 
%--
which is the source's line-of-sight direction, we have $\cos \theta =
\hat{\boldsymbol N} \cdot \hat{\boldsymbol z}$, $\cos \iota = \hat{\boldsymbol
N} \cdot \hat{\boldsymbol L}$, and
%--
\begin{align}
	\phi &= \arctan \left(\frac{\hat{\boldsymbol N} \cdot \hat{\boldsymbol y}}{\hat{\boldsymbol N} \cdot
	\hat{\boldsymbol x}}\right) +  \alpha_0\,,\\
	\tan \psi &=  \frac{\hat{{\boldsymbol L}} \cdot \hat{{\boldsymbol z}}-(\hat{{\boldsymbol L}} \cdot \hat{{\boldsymbol N}})(\hat{{\boldsymbol z}} \cdot \hat{{\boldsymbol N}})}{\hat{{\boldsymbol N}} \cdot(\hat{{\boldsymbol L}} \times \hat{{\boldsymbol z}})}\,.
\end{align}
%--
Plugging them into Eqs.~(\ref{eq:Fplus}--\ref{eq:Fcross}), we finally derived
the ground-based pattern functions.

The Doppler phase, $\varphi_{D}(t)={2 \pi f} R_E  \cos \theta$, contains
the information of the time required for the waves to travel from the geocenter
to reach the detector \citep{Zhao:2017cbb}, where $R_E$ is the radius of the
Earth. We ignore the Doppler effect due to the Earth's motion around the Sun in
the calculation. It turns out that such omission does not affect the
localization precision.

In addition, by transforming the ecliptic coordinate to geocentric coordinate,
then substituting it into the PyCBC \citep{pycbc} pattern function code, we have
cross-checked the validity of our method. Note that in the above we only model
the response of rectangular detectors such as the CE. For triangular ones, one needs
to multiply $\sqrt3/2$ to $F^+$ and $F^\times$. We have now obtained $F^+$,
$F^\times$, and $\varphi_D$ for both space and ground detectors. 

\begin{figure}%[htb!]
	\centering
	\includegraphics[width=\linewidth]{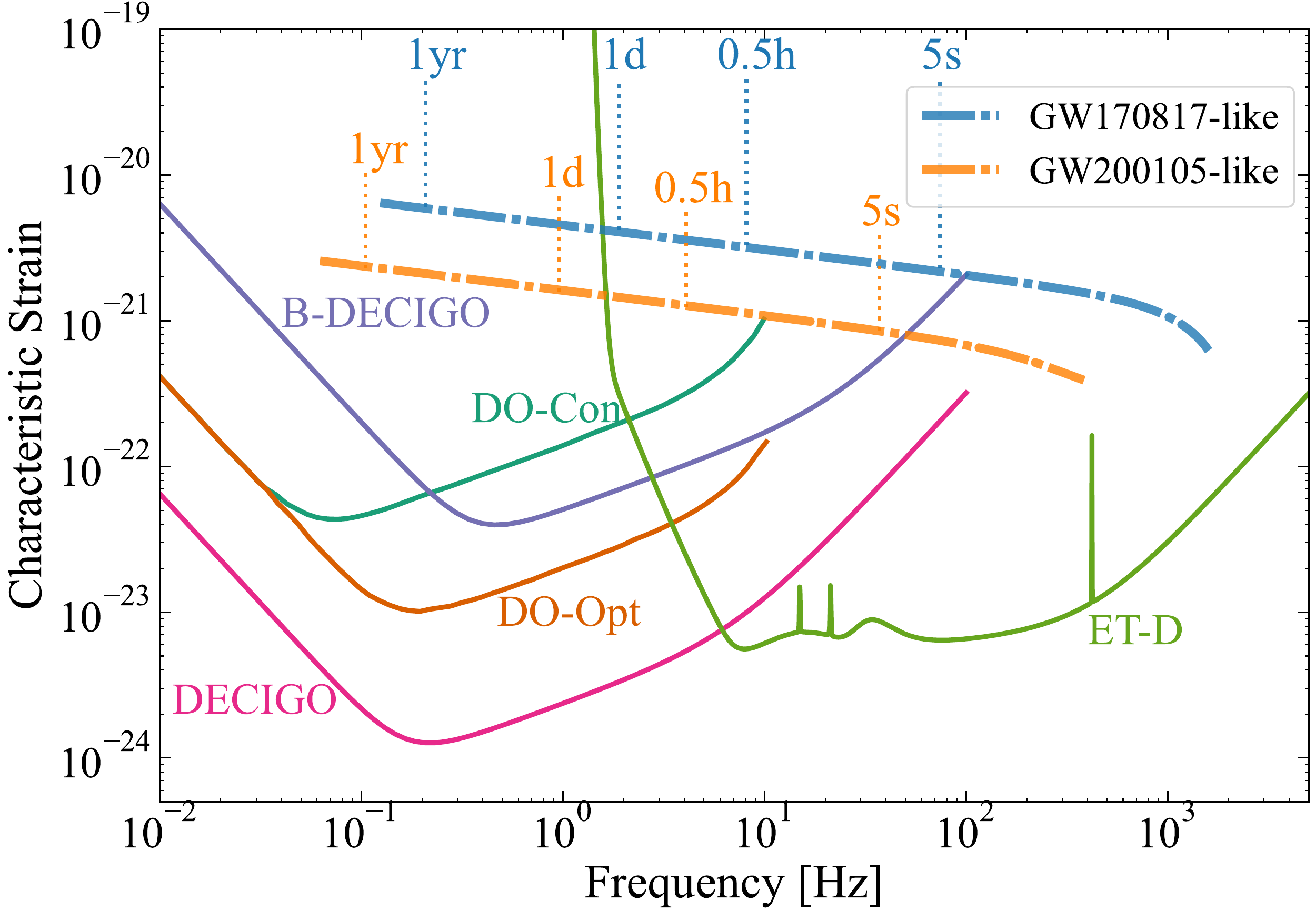}
	\caption{The strain amplitude of the example sources, ${2f| \tilde
	h_+(f)|}$, and the characteristic strain of the detector noise,
	$\sqrt{fS_{n}^{\rm eff}(f)}$, where $S_{n}^{\rm eff}(f)$ is the sky-averaged
	effective noise defined in Eq.~(29) of \citet{Liu:2020nwz}. Source signals
	are plotted for a duration of $T_{\rm obs} = 4$\,yrs. For each source,
	dashed lines mark the times before coalescence.} \label{fig:strain}
\end{figure}

To explore multiband enhancement, for the decihertz observatories, we choose
four designs,  namely B-DECIGO, DECIGO, DO-Conservative, and DO-Optimal, and we
use ET as a representative of the hectohertz ground-based detector. Below we
give details on the equivalent number of detectors, the geometrical
configuration, the frequency ranges and the relevant literature to obtain their
noise power spectral density (PSD). 
%--
\begin{itemize}
	\item For DO-Optimal and DO-Conservative, we use two effective detectors,
	and triangular LISA-like orbits. Their  sensitivity curves are taken from
	\citet{Sedda:2019uro}, which we treat as the averaged PSD over $\theta$,
	$\phi$, $\psi$, and detector numbers. The frequency range is
	$[f_{\rm{low}},f_{\rm{high}}] = [10^{-3},10]$\,Hz.
	\item For B-DECIGO, we use two effective detectors, and a triangular LISA-like
	orbit. The sensitivity curve is taken from Eq.~(20) of
	\citet{Isoyama:2018rjb}, and the frequency range is
	$[f_{\rm{low}},f_{\rm{high}}] = [10^{-2},100]$\,Hz.
	\item For DECIGO, we use eight effective detectors with four triangular
	LISA-like interferometers located from one another by 120$^\circ$ separation
	on their heliocentric orbits. The sensitivity curve is taken from Eq.~(5) of
	\citet{Yagi:2011wg}, and the frequency range is $[f_{\rm{low}},f_{\rm{high}}]
	= [10^{-3},100]$\,Hz.
	\item For ET, we adopt the final design ET-D, which has three triangular
	detectors and possibly be placed at Italy; so we set the latitude of ET
	$\delta=0.7615$. The sensitivity curve is taken from \citet{Hild:2010id},
	and the frequency range is $[f_{\rm{low}},f_{\rm{high}}] = [1,10^4]$\,Hz.
\end{itemize}
%--
The sky-averaged noise curves of these GW detectors are given in Fig.~\ref{fig:strain}.

Throughout the paper, we mainly study the PE using synergy of B-DECIGO and ET as
a fiducial scenario.  We will make comparison with the other three space-borne
detectors specifically in Sec.~\ref{sec:other_det}. 

%---------------------------------------------------------------------
%---------------------------------------------------------------------

%---------------------------------------------------------------------
\subsection{PE Method and Source Selection}
\label{sec:pe method}
%---------------------------------------------------------------------

We use matched filtering to estimate the binary parameters \citep{Finn:1992wt,
Cutler:1994ys}. The noise weighted inner product between two signals, $h(t)$ and
$g(t)$, is defined as 
%--
\begin{equation}
	(h|g) \equiv 2 \int_{f_{\rm in}}^{f_{\rm out}} \frac{
	{\tilde{h}^* (f) } \tilde{g} (f) + {\tilde{g}}^* (f)
	\tilde{h} (f) } {S_n(f)} {\rm d} f \,, \label{eq:innerproduct}
\end{equation}
%--
where $ {S_n(f)}$ is the noise PSD of the detector; the frequency range ${f_{\rm
in}}$ and ${f_{\rm out}}$ are determined by the detector's limitation and the
property of the signal by ${f_{\rm in}}={\rm max}(f_{\rm{4\,yr}},f_{\rm low})$
and ${f_{\rm out}}={\rm min}(f_{\rm{ISCO}},f_{\rm high})$, where $f_{\rm 4\,yr}
= \left({ t_{\rm obs}}/{5}\right)^{-3/8}{\cal M}^{-5/8} / 8 \pi$ with $t_{\rm
obs}=4\,\rm{yr}$ is the GW frequency 4 years before the merger, and $f_{\rm
ISCO}=\left(6^{3/2} \pi M \right)^{-1}$ is the the GW frequency at the innermost
stable circular orbit (ISCO) of a Schwarzschild metric with mass $M$. We list
$f_{\rm in}$ and $f_{\rm out}$ for different sources in Table~\ref{tab:source}.

The SNR for a signal $h$ is given by $\rho \equiv \sqrt{(h | h)}$.  In the limit
of large SNRs, supposing that the noise is stationary and Gaussian, the Fisher
matrix method \citep{Finn:1992wt} is a fast way to estimate parameter
statistical errors.  We denote a collection of parameters in a vector,
$\bm{\Xi}$. The element of the Fisher matrix $\Gamma_{ab}$ is then given by 
$\Gamma_{ab} \equiv \left( {\partial h}/{\partial \Xi^a} \right| \left.
{\partial h}/{\partial\Xi^b} \right)$, where $h$ is the detector-frame GW
strain, i.e. Eq.~(\ref{eq:det_response}).  The error vector, ${\Delta
\bm{\Xi}}$, has a multi-variate Gaussian probability distribution
\citep{Vallisneri:2007ev}, $p(\Delta{\bm{\Xi}})\propto\exp \left(- \Gamma_{a b}
\Delta \Xi^{a} \Delta \Xi^{b} / 2\right)$, where $\Delta \Xi^{a} \equiv \Xi^{a}
- \hat{\Xi}^{a}$ with $\hat{\Xi}^{a}$ the maximum-likelihood parameter
determined by the matched filtering.  The variance-covariance matrix element is
given by $\left\langle\delta \Xi^{a} \delta
\Xi^{b}\right\rangle=\left(\Gamma^{-1}\right)^{ ab}$, then an estimate of the
root-mean-square (rms), $\Delta \Xi^{a}$,  and the cross-correlation between $
\Xi^{a}$ and $ \Xi^{b}$, $c^{ab}$, are $\Delta \Xi^{a} =\sqrt{
\left(\Gamma^{-1}\right)^{a a}}$ and $ c^{ab} = {\left\langle\delta \Xi^{a}
\delta \Xi^{b}\right\rangle}/{\Delta \Xi^{a}\Delta \Xi^{b}}$, respectively.  The
angular resolution $\Delta \Omega$ is defined as $\Delta \Omega=2 \pi
\big[\left(\Delta \bar\mu_{S} \Delta \bar\phi_{S}\right)^{2}-\left\langle\delta
\bar\mu_{S} \delta \bar\phi_{S}\right\rangle^{2}\big]^{1/2}$, where $\bar\mu_{S}
\equiv \cos\bar\theta_S$ \citep{Lang:2007ge, Barack:2003fp}.  Finally, to
estimate parameter precision from joint observations, we add the Fisher matrices
from both detectors together via, $\Gamma_{ab}^{\mathrm{joint}}
=\Gamma_{ab}^{\mathrm{space}}+\Gamma_{ab}^{\mathrm{ground}}$
\citep{Cutler:1994ys}.
  
%---------------------------------------------------------------------
\begin{table}%[htb!]
\caption{Properties of the GW170817-like BNS system and the GW200105-like NSBH
system explored in the paper. We choose a fixed angular momentum direction for
both sources for the convenience of later comparison. The fiducial value of
$\Lambda_1$, $\Lambda_2$, $\kappa_1$, and $\kappa_2$ are set such that their
values represent typical NSs and BHs, and the values of $\tilde Q$,
$\delta\tilde Q$, $\tilde\Lambda$, and $\delta\tilde\Lambda$ are derived using
equations in Sec.~\ref{sec:wf}.}
\label{tab:source}
\centering
\begin{tabular}{lll}
  \hline\hline
   & {GW170817-like}  & {GW200105-like}  \\
  \hline 
  $m_1$ ($M_{\odot}$) & {$1.46$} & {$8.9$}  \\
  $m_2$ ($M_{\odot}$) & {$1.27$} & {$1.9$}  \\
  $\chi_1$ & {$0.0469$} & {$0.125$}  \\
  $\chi_2$ & {$0.002$} & {$0.004$}  \\
  $\kappa_1$ & {$9$} &  {$1$} \\
  $\kappa_2$ & {$10$} &  {$3$} \\
  $\Lambda_1$ & {$675$} &  {$ 0$} \\
  $\Lambda_2$ & {$951$} &  {$ 237$} \\
  $\tilde\Lambda$ & {$793$} & {$2.81$} \\ 
  $\tilde Q$   & {$1.01\times10^{-2}$} & {$1.98\times10^{-6}$} \\
  $\cos\bar \theta_L$ & {$-0.65$} & {$-0.65$} \\
  $\bar \phi_L$ (${\rm rad}$) & {$5.016$} & {$5.016$}\\
  $D_L$ (Mpc)  & {$40$} & {$280$} \\
  $z$  & {$0.01$} & {$0.06$} \\
  $t_{\rm obs, DECIGO/DO}$   &  4\,yr & 4\,yr \\ 
  $t_{\rm obs, ET}$ & 5.6\,d & 0.90\,d \\ 
  $f_{\rm in, DECIGO/DO}$ (Hz) &  0.124 & 0.0622  \\ 
  $f_{\rm in, ET}$ (Hz) &1.0 & 1.0 \\ 
  $f_{\rm out, DECIGO}$ (Hz) &  100 & 100 \\
  $f_{\rm out, DO}$ (Hz) &  10 & 10 \\
  $f_{\rm out, ET}$ (Hz) &  1595 & 384.1 \\
\hline 
\end{tabular}
\end{table}
%---------------------------------------------------------------------

Now we turn to source selection. Because we are interested in both BNS and NSBH
systems, we choose our fiducial values from the properties of (i) the BNS
inspiral GW170817, and (ii) the NSBH merger GW200105. Meanwhile, we take
reasonable values for the poorly measured parameters such as $\chi_{1,2}$,
$\kappa_{1,2}$, and $\Lambda_{1,2}$. We list source properties in
Table~\ref{tab:source}. Furthermore, we also select three fixed locations for
later comparisons: (I) $\cos \bar \theta_L = 0$ and $\bar \phi_S = 2.0$, (II)
$\cos \bar \theta_L = 0.271$ and $\bar \phi_S = 0$, and (III) $\cos \bar
\theta_L = 0.936$ and $\bar \phi_S = 4.768$.  We will refer to the BNS system at
location I/II/III as ``BNS I/II/III'' and the NSBH analog as ``NSBH I/II/III''
in the following analyses.  As we will see, location I has a large SNR and
location III has a precise sky localization.

Finally, we define three parameter sets for the convenience of explication: (i)
the intrinsic parameter set,
%--
\begin{equation}
	\bm{\Xi}^{\rm int} = \big\{  {\cal M}, \eta,	 \chi_s, \tilde Q,\tilde\Lambda
	\big\} \,;
\end{equation}
%--
(ii) the extrinsic parameter set
%--
\begin{equation} \label{eq:ext}
	\bm{\Xi}^{\rm ext} \equiv \big\{ t_c, \phi_c, D_{L}, 
\bar \theta_S, \bar\phi_S,\bar\theta_L,
	\bar\phi_L\big\} \,;
\end{equation}
%--
and (iii) the localization parameter set which is a subset of $\bm{\Xi}^{\rm
ext}$,
%--
\begin{equation} \label{eq:loc}
	\bm{\Xi}^{\rm loc} \equiv \big\{\bar \theta_S, \bar\phi_S,\bar\theta_L,
	\bar\phi_L, D_{L}\big\}  \subset \bm{\Xi}^{\rm ext} \,.
\end{equation}
%--
As a short summary, the parameters that we put into the waveforms are,
%--
\begin{equation} \label{eq:Input}
	\bm{\Xi}^{\rm input} \equiv \big\{m_1,m_2,\chi_1,\chi_2,\kappa_1,\kappa_2,\Lambda_1,\Lambda_2 \big\}  \cup \bm{\Xi}^{\rm ext} \,,
\end{equation}
%--
whereas the parameters we estimate are, 
%--
\begin{equation} 
\label{eq:PE}
	\bm{\Xi}^{\rm PE} \equiv  \bm{\Xi}^{\rm int} \cup \bm{\Xi}^{\rm ext} \,.
\end{equation}
%--

For the spin parameters, we choose only to estimate $\chi_s$ mainly for two
reasons: (i) when simultaneously estimating $\chi_a$ and $\chi_s$, or $\chi_1$,
$\chi_2$, the correlations between them, as well as with $\tilde Q$, become
larger than 0.9999 such that the Fisher matrix will be rather singular, while
estimating $\chi_s$ is slightly uncorrelated than estimating $\chi_1$, $\chi_2$,
or $\chi_a$; (ii) from the formation channel point of view, a BNS system often
consists of a rapidly spinning, recycled pulsar and a slowly rotating,
second-born pulsar whose $\chi$ is very close to zero \citep{Tauris:2017omb}, so
estimating one of the spin parameter is sufficient to constrain such a system
within an astrophysical setting for field binaries.

\rep{It is worth noting that when the contribution of $\delta\tilde\Lambda$
grows, the omission of $\delta\tilde\Lambda$ in the estimation could lead to
over-estimated constraints on $\Delta\tilde\Lambda$. On the other hand, the lack
of the prior knowledge in our consideration could under-estimate the parameter
errors. Quantitatively, we have checked that both kinds of effects on the
uncertainties are less than one order of magnitude.}

In calculating the Fisher matrix, the analytical expressions for the partial
 derivative ${\partial \tilde{h}}/{\partial \Xi^a}$ are usually not available.
 We decide to calculate the partial derivatives of $\tilde{h}(f)$ with respect
 to $t_c$, $\phi_c$, $\tilde Q$, and $\tilde\Lambda$ analytically, and calculate
 the partial derivatives of $\tilde{h}(f)$ with respect to the rest parameters
 numerically.  For the latter, we adopt a numerical scheme that ${\partial
 \tilde{h}}/{\partial \Xi^a}=\tilde [h( \Xi^a+\delta_{\Xi^a})-h(
 \Xi^a-\delta_{\Xi^a})]/2\delta_{\Xi^a}$, and we have chosen $\delta_{\Xi^a}$
 for each parameter carefully such that the PE results are stable.

%---------------------------------------------------------------------
\begin{deluxetable*}{lllDDDcccDDl}
\tablecaption{The parameter uncertainties of some typical systems. The first two
columns label the source type, location and the detectors used.  The table is
vertically divided into 3 parts, (i) the first part illustrates the effect of
different locations for BNSs I, II, and III; (ii) the second part illustrates
the effect of different detectors for BNS II; and (iii) the third part
illustrates results for NSBH II for comparison with the corresponding BNS
results.
\label{tab:result}
}
\tablewidth{0pt}
\tablehead{
 \colhead{System}& \colhead{Detector} & \colhead{SNR} & \multicolumn2c{$\Delta {\cal M}/{\cal M}$} 
&\multicolumn2c{$\Delta\eta/\eta$} & \multicolumn2c{$\Delta \chi_s/\chi_s$} & \colhead{$\Delta\tilde Q/\tilde Q$} 
& \colhead{$\Delta\tilde\Lambda/\tilde\Lambda$}  & \colhead{$\Delta t_c$} & \multicolumn2c{$\Delta \phi_c$} 
& \multicolumn2c{$\Delta D_L/D_L$} & \colhead{$\Delta \Omega$} 
\\
 \colhead{} & \nocolhead{} & \colhead{} & \multicolumn2c{$(10^{-6})$}
& \multicolumn2c{$(10^{-3})$} & \multicolumn2c{$(10^{-2})$} & \colhead{} & & \colhead{(ms)} & \multicolumn2c{$(10^{-1})$} & \multicolumn2c{$(10^{-2})$} & \colhead{$({\rm arcmin}^2)$}
}
\decimalcolnumbers
\startdata
 BNS I & ET &  1340 & 0.58 & 1.5 & 0.64 & 1.4 & 0.0053 & 0.13 & 0.11 & 0.71 & 2.55\,(deg$^2$) \\
 & B-DEC &  212 & 0.023 & 0.59 & 4.5 & 5.6 & 27 & 12 & 5.4 & 3.3 & 9.08 \\
 & B-DEC+ET &  1360 & 0.0037 & 0.050 & 0.18 & 0.16 & 0.0040 & 0.040 & 0.10 & 0.49 & 0.431 \\
 \midrule[0.05mm]
 BNS II & ET &  288 & 1.9 & 5.2 & 2.8 & 5.2 & 0.024 & 0.26 & 0.23 & 0.96 & 0.803\,(deg$^2$) \\
  & B-DEC &  99.6 & 0.051 & 1.2 & 8.6 & 10 & 42 & 20 & 9.2 & 1.0 & 0.309 \\
  & B-DEC+ET &  305 & 0.0092 & 0.13 & 0.55 & 0.49 & 0.017 & 0.028 & 0.14 & 0.45 & 0.0128 \\
   \midrule[0.05mm]
  BNS III & ET &  374 & 1.6 & 4.4 & 2.3 & 4.3 & 0.019 & 0.068 & 0.20 & 1.2 & 0.0684\,(deg$^2$) \\
  & B-DEC &  95.2 & 0.050 & 1.2 & 8.5 & 10 & 42 & 20 & 9.3 & 1.7 & 0.0797 \\
  & B-DEC+ET &  386 & 0.0088 & 0.13 & 0.51 & 0.44 & 0.014 & 0.023 & 0.13 & 0.57 & 0.00320 \\
\midrule[0.2mm]
 BNS II 
  & DO-Con &  67.4 & 0.15 & 5.2 & 50 & 72 & 3300 & 300 & 96 & 1.5 & 0.432 \\
  & DO-Con+ET &  296 & 0.011 & 0.17 & 0.71 & 0.61 & 0.018 & 0.030 & 0.15 & 0.49 & 0.0837 \\
   \midrule[0.05mm]
  & DO-Opt &  406 & 0.025 & 0.82 & 7.9 & 11 & 470 & 44 & 15 & 0.25 & 0.0118 \\
  & DO-Opt+ET &  498 & 0.0029 & 0.053 & 0.28 & 0.28 & 0.016 & 0.025 & 0.13 & 0.21 & 0.00249 \\
   \midrule[0.05mm]
  & DECIGO &  3240 & 0.0014 & 0.040 & 0.33 & 0.43 & 4.1 & 0.77 & 0.47 & 0.031 & 7.79$\times 10^{-6}$ \\
  & DECIGO+ET &  3260 & 0.00060 & 0.013 & 0.087 & 0.097 & 0.013 & 0.017 & 0.070 & 0.031 & 7.68$\times 10^{-6}$ \\
\midrule[0.2mm]
 NSBH II& ET &  105 & 54 & 48 & 5.7 & $>10^5$
 %1.6e+05 
 & 8.7 & 3.8 & 3.5 & 13 & 184\,(deg$^2$) \\
 & B-DEC &  36.9 & 0.3 & 2.6 & 5.7 & $> 10^4$
% 7.4e+04 
 & 1400 & 170 & 11 & 2.7 & 32.8 \\
 & B-DEC+ET &  112 & 0.079 & 0.42 & 0.56 & $>10^3$
% 6.1$\times 10^{3}$ 
 & 3.3 & 0.19 & 0.29 & 1.2 & 0.354 \\
 \hline
 \enddata
\tablecomments{When only ET is used, $\Delta\Omega$ is given in the unit of square degrees.}
\end{deluxetable*}
%---------------------------------------------------------------------

%---------------------------------------------------------------------
\begin{figure}
\includegraphics[width=1.05\linewidth]{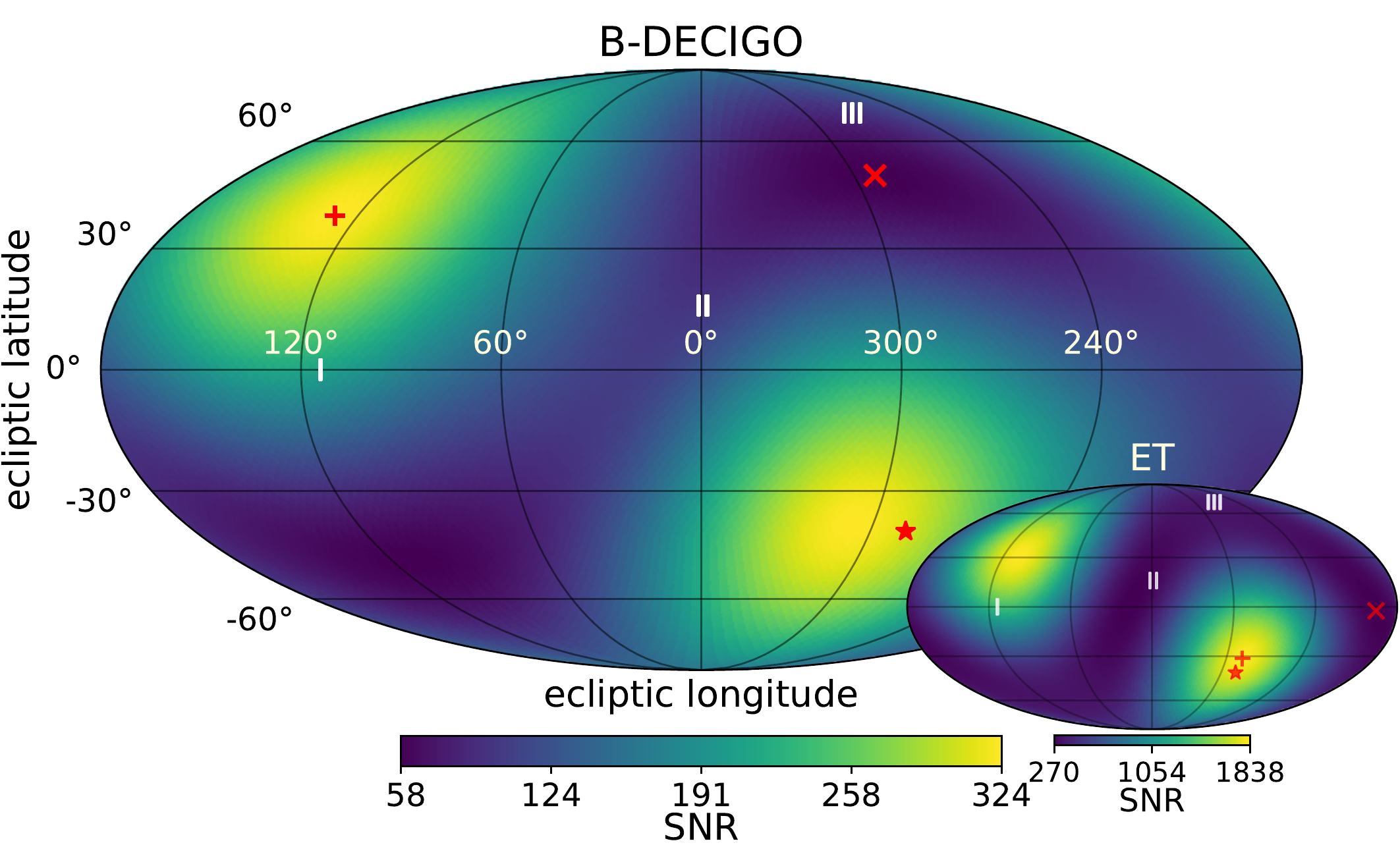}
\caption{The SNR sky map of BNSs for B-DECIGO (large) and ET (small)
in ecliptic coordinate. The red star marks the direction of
$\hat{\boldsymbol{L}}$; the red ``$\times$'' (``$+$'') marks the lowest
(highest) SNR in each map. The Roman numbers I, II, and III, mark the source
locations I, II, and III. Note that we have used HEALPix\textsuperscript{a} to
help produce the sky maps and the resolution of the map is $\approx 3.4\,{\rm
deg^2}$. We plot in ecliptic coordinate for all the sky maps in this paper.}
\small\textsuperscript{a} \url{http://healpix.sf.net}
\label{fig:SNR}
\end{figure}
%---------------------------------------------------------------------

%---------------------------------------------------------------------
\section{Results}
\label{sec:result}
%---------------------------------------------------------------------

In this section, we present our PE results focusing on the constraints of
quadrupole parameter, tidal deformability and the localization precision. 

We show the dependence of SNR on the source sky position for B-DECIGO and ET
respectively in the large and small sky maps in Fig.~\ref{fig:SNR}. For the
convenience of reading Fig.~\ref{fig:SNR}, as well as the following
Figs.~\ref{fig:Q}--\ref{fig:loc_B}, and Fig.~\ref{fig:other_det}, we explain the
common characteristics of such sky maps here. The plots are based on ecliptic
coordinate, showing the signal SNR or PE precision as a function of source's sky
location.  The red star in each sky map marks the direction of the source's
angular momentum $\hat{\boldsymbol{L}}$, and the red ``$\times/+$'' marks the
lowest/highest value in each map. The Roman numbers, I, II, and III, mark the
source locations I, II, and III. 

From Fig.~\ref{fig:SNR} we see that the SNR is larger when the source is either
face on or face off, which is intuitive. The slight deviation of the maximum
point location is caused by the detector's orbit. Meanwhile, the minimum point
is perpendicular to this direction. In later analyses we show that SNR is not
the dominant factor to the PE precision, especially for localization. 

We also tabulate the complete PE results in Table~\ref{tab:result} for the
example sources whose properties are listed in Table~\ref{tab:source}. It shows
the comparison of PE results between different sky locations, different
detectors and different systems. We will analyze them in detail in the following
subsections.

In Sec.~\ref{sec:corr}, we present the parameter correlations from single/joint
detection for BNS/NSBH systems.  Based on different features of $\bm{\Xi}^{\rm
int}$ and $\bm{\Xi}^{\rm ext}$, we analyze the constraints on them respectively
in Sec.~\ref{sec:int} and Sec.~\ref{sec:ext}. More specifically, in
Sec.~\ref{sec:int}, we show the multiband improvements on $\bm{\Xi}^{\rm int}$,
with Sec.~\ref{subsec:Q} focusing on $\Delta\tilde Q$ and Sec.~\ref{subsec:L}
focusing on $\Delta\tilde\Lambda$, and discuss different characteristics of
BNS/NSBH systems. In Sec.~\ref{sec:ext}, we show the features of $\bm{\Xi}^{\rm
ext}$, in which Sec.~\ref{sec:Omega} investigates the localization ability from
space, ground, and multiband observations, and Sec.~\ref{sec:alert} displays the
evolution of angular resolution with the passing of observing time, which
provides information for early warning alerts. In Sec.~\ref{sec:other_det}, we
analyze the different constraints of using DOs/B-DECIGO/DECIGO jointly with ET
and in Sec.~\ref{sec:PV2}, we briefly compare our results with the PE result
using another phenomenological waveform, i.e.
\texttt{IMRPhenomPv2\char`_NRTidalv2}.

%---------------------------------------------------------------------
\begin{figure*}
\centering
\includegraphics[width=0.7\linewidth]{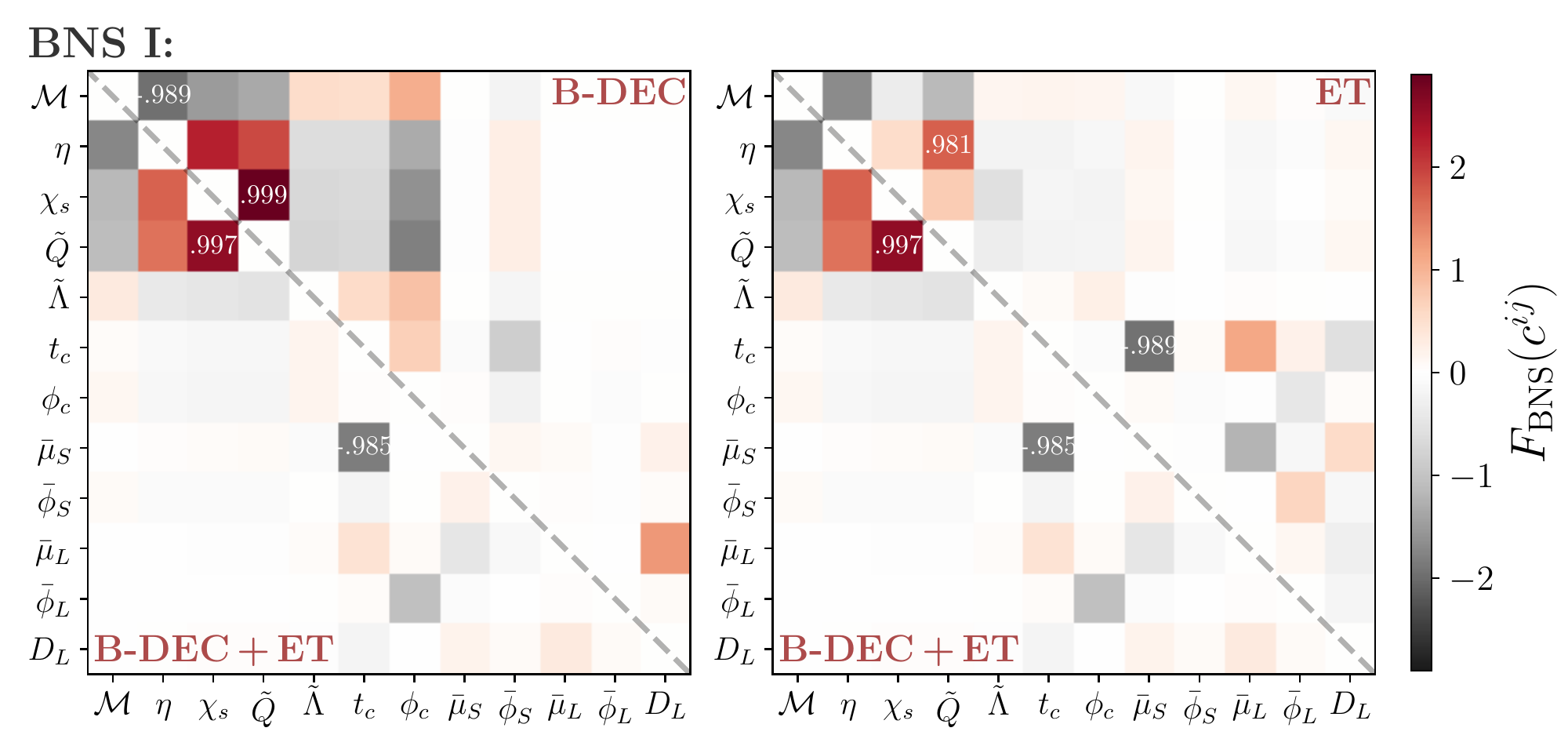}
\includegraphics[width=0.7\linewidth]{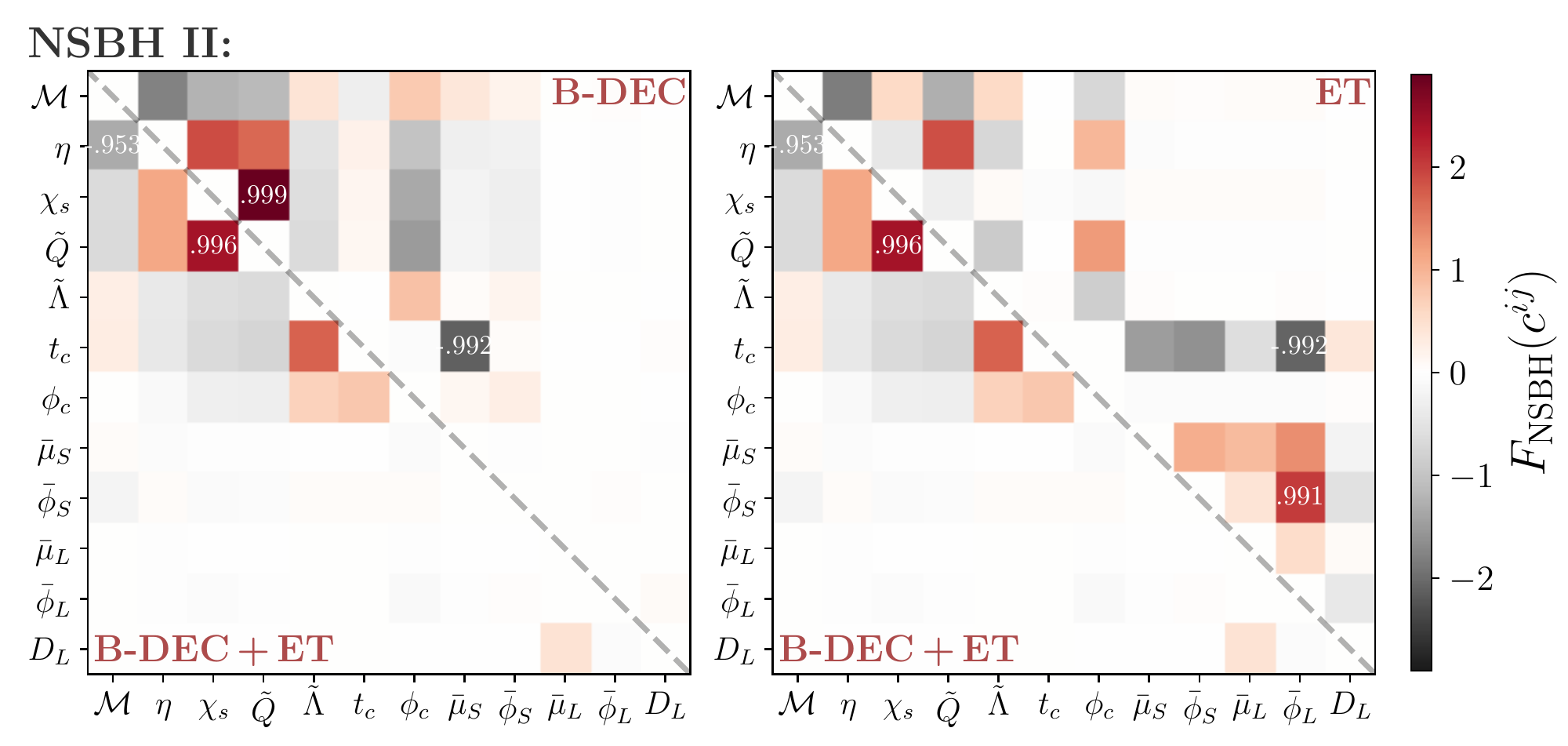}
\caption{The correlation matrices from the Fisher matrix analysis. The upper
(lower) two panels are the parameter correlations for BNS I (NSBH II).  In each
panel, the upper triangular matrix is for single detectors (B-DEC in the left
column and ET in the right column), and the lower triangular matrix is for joint
detection (B-DEC+ET).  We indicate the name of the detectors in  brown in the
corresponding corners.  The color uses a function, $F(c^{ij})\equiv
\log_{10}[(1+c^{ij})/(1-c^{ij})] - c^{ij}\log_{10}2$, adopted from
\citet{Shao:2016ubu}, which is  defined such that it counts the number of $9$'s
after the decimal point in the cases of large correlations. The largest positive
and negative corrections are explicitly given in values for each triangle
matrix.}\label{fig:corr}
\end{figure*}
%---------------------------------------------------------------------

%---------------------------------------------------------------------
\subsection{Parameter Correlations}
\label{sec:corr}
%---------------------------------------------------------------------

Figure~\ref{fig:corr} presents with a specific design the parameter correlations
in systems BNS I (upper panels) and NSBH II (lower panels), each with three
observing schemes (ET, B-DECIGO, and ``B-DECIGO+ET''). We show the correlations
between all 12 parameters in $\bm{\Xi}^{\rm PE}$ observed by single detectors in
the upper right corners of four panels, with B-DECIGO in the left two panels and
ET in the right two panels.  We show the joint detection of ``B-DECIGO+ET'' in
the lower left corners in all of four panels with repetition for left and right
columns. The shades of the color indicate the degrees of correlation.
Intuitively, we see that for both kinds of sources, the boundary between
intrinsic parameters $\bm{\Xi}^{\rm int}$ and location parameters $\bm{\Xi}^{\rm
loc}$ is evident. The other extrinsic parameters, such as $t_c$ and $\phi_c$,
weakly correlate with both $\bm{\Xi}^{\rm int}$ and $\bm{\Xi}^{\rm loc}$. 

Relating the correlation with the PE results in Table~\ref{tab:result}, we find
that the degree of correlation between parameters to some extent indicates the
degree of parameter precision one could limit.  (i) Space-borne detectors are
good at localization, therefore the correlation among the parameters in
$\bm{\Xi}^{\rm loc}$ for B-DECIGO is small; see the lower right zone of the
B-DECIGO triangular matrix in Fig.~\ref{fig:corr}. ET, on the other hand, has a
larger correlation, such as 0.991 between $\bar \phi_S$ and $\bar \phi_L$ for
the NSBH II system.  From the red and black colors of the lower right of two ET
triangular matrices we notice that the correlation among the parameters in
$\bm{\Xi}^{\rm loc}$ is more obvious in the NSBH II system, since the duration
of the NSBH signal in ET is too short to achieve a good localization.  (ii)
Ground-based detectors are good at constraining intrinsic parameters that enter
at high PN orders. This is natural because the high PN terms contribute
information largely at the higher frequency band. At the same time, the
correlation among parameters in $\bm{\Xi}^{\rm int}$ is smaller than that of
B-DECIGO, especially between $\chi_s$, $\tilde Q$, and $\tilde \Lambda$.  The
correlation between parameters in $\bm{\Xi}^{\rm int}$ and $\bm{\Xi}^{\rm loc}$
is also  small for this reason; see the upper rectangle zone of the ET matrices.

In each panel, by comparing the upper triangular matrix with the lower
triangular matrix we could see the multiband enhancement. B-DECIGO has large
degeneracy between $\chi_s$ and $\tilde Q$, with $c^{\chi_s \tilde Q}$ reaching
0.999 in both BNS I and NSBH II systems. Joint detection could lower the
correlation to $c^{\chi_s \tilde Q}=0.997$ for BNS I and $c^{\chi_s \tilde
Q}=0.996$ for NSBH II. ET has large degeneracy among parameters in $\bm{\Xi}^{\rm
ext}$, while joint detection could largely reduce this degeneracy. It could even
reduce the correlation to nearly zero for the NSBH system.

In Fig.~\ref{fig:corr}, we only show the correlation matrices at locations I and
II. However, for sources whose line of sight direction is aligned or
anti-aligned with its angular momentum direction, the correlations between
$\phi_c$ and $\mu_L$, as well as between $D_L$ and $\phi_L$, become larger. They
might reach up to 0.9999 and could get even worse in multiband observations.
Later we will see that, the red triangles in Fig.~\ref{fig:loc_B} and
Fig.~\ref{fig:other_det} mark the locations where the maximum correlation of the
matrix is larger than 0.9995. For sources in these places, the multiband
observations cannot break the degeneracy between parameters in $\bm{\Xi}^{\rm
loc}$ at all. Nevertheless, this only happens for these particularly unfavoured,
sky locations. 

Based on the above arguments, considering the difference between intrinsic and
extrinsic parameters, we will analyze them separately in the following two
subsections.

%---------------------------------------------------------------------
\subsection{Estimation on Intrinsic Parameters}
\label{sec:int}
%---------------------------------------------------------------------

In this subsection, we illustrate how multiband observations improve intrinsic
parameters in $\bm{\Xi}^{\rm int}$, especially focusing on the quadrupole
parameter $\tilde Q$ and the tidal deformability $\tilde \Lambda$. Before
digging into them respectively in Sec.~\ref{subsec:Q} and Sec.~\ref{subsec:L},
we present some general characteristics first.

Columns (4--8) in Table~\ref{tab:result} show the precision of parameters in
$\bm{\Xi}^{\rm int}$ for the selected sources. From the first part of the table
we find that with the change of location, the parameter precision is
approximately inversely proportional to the SNR. But unlike the relation between SNR
and distance $D_L$, this inverse relation is not exact. For example, using
B-DECIGO (ET), the SNR is 2.13 (4.65) times higher at location I than that of
location II, but the parameter precision improvement for $\{\Delta {\cal M},
\Delta\eta,	 \Delta\chi_s, \Delta\tilde Q, \Delta\tilde\Lambda \}$ is \{2.17,
2.02, 1.90, 1.84, 1.55\} (\{3.29, 3.56, 4.31, 3.81, 4.47\}) times instead. The
deviation of such scaling with SNR indicates the impact of source direction and
orientation on estimating parameters in $\bm{\Xi}^{\rm int}$.

During the investigation, we notice that the change of the fiducial values for
$\tilde{Q}$ and $\tilde{\Lambda}$ within reasonable ranges will not
affect the PE results significantly. In general, with the increasing  fiducial
values, the estimated errors decrease slightly within an order of magnitude.

\begin{figure}
\gridline{\fig{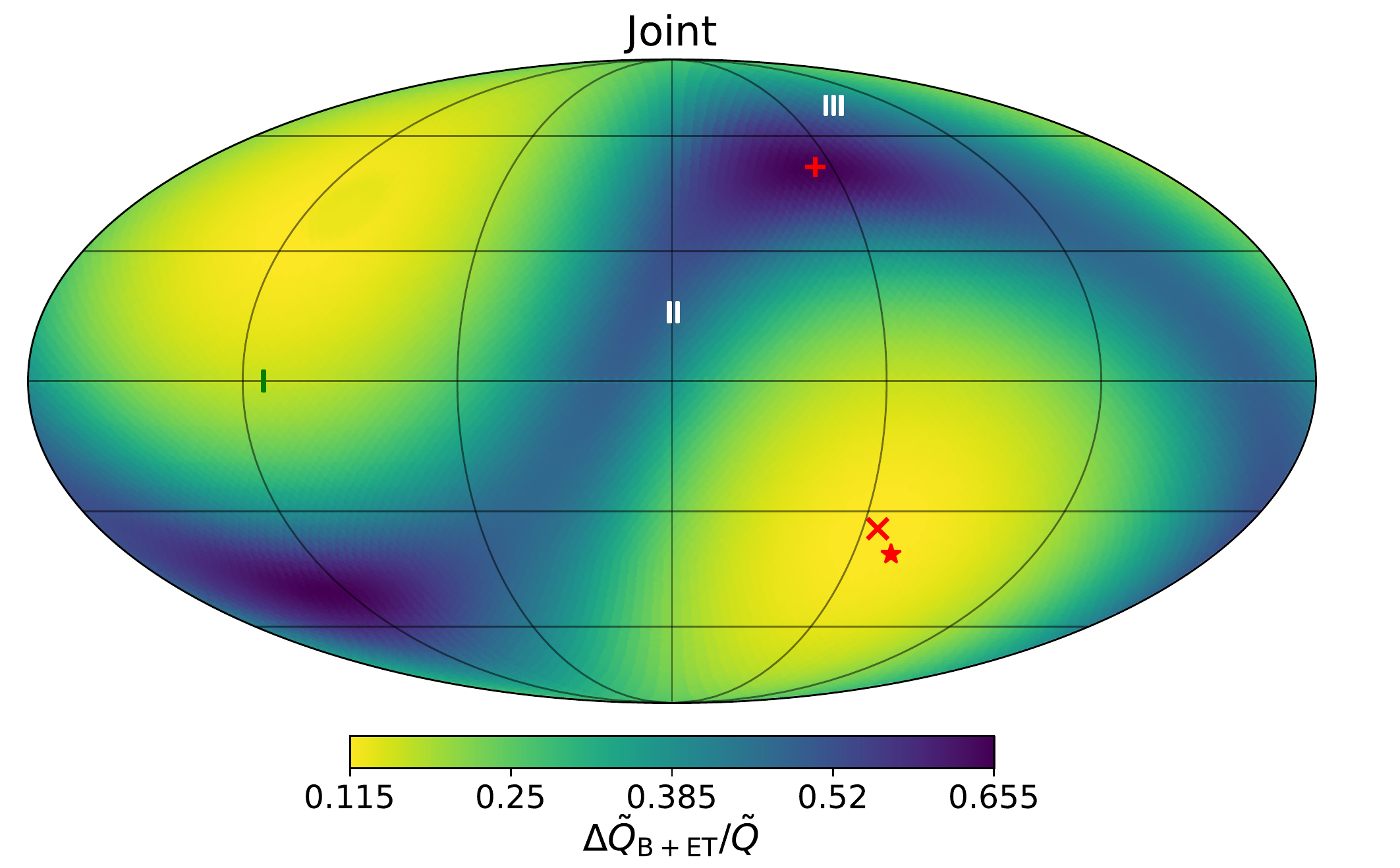}{0.5\textwidth}{(a)}}
\vspace{-0.4cm}
\gridline{\fig{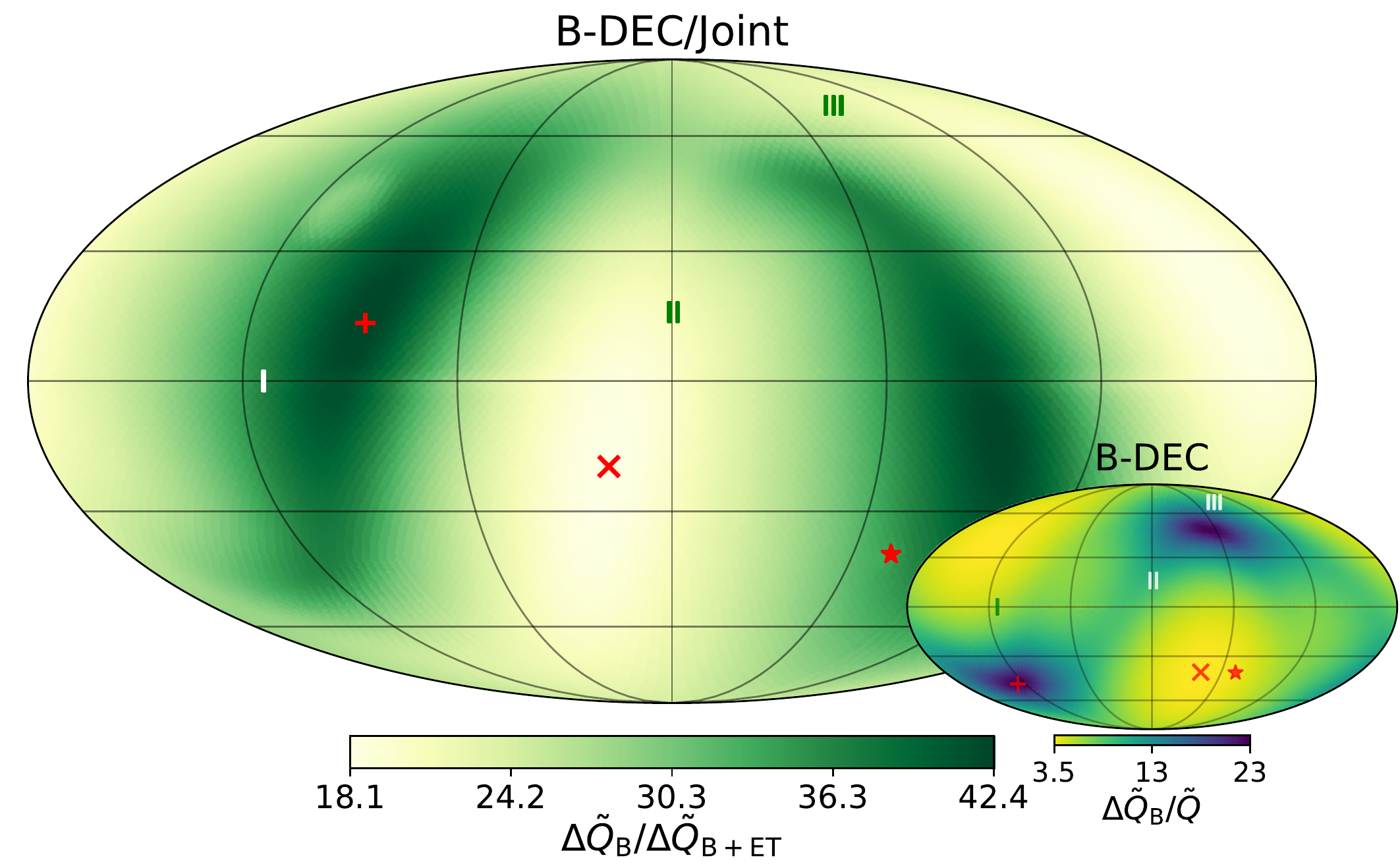}{0.5\textwidth}{(b)}}
\vspace{-0.4cm}
\gridline{\fig{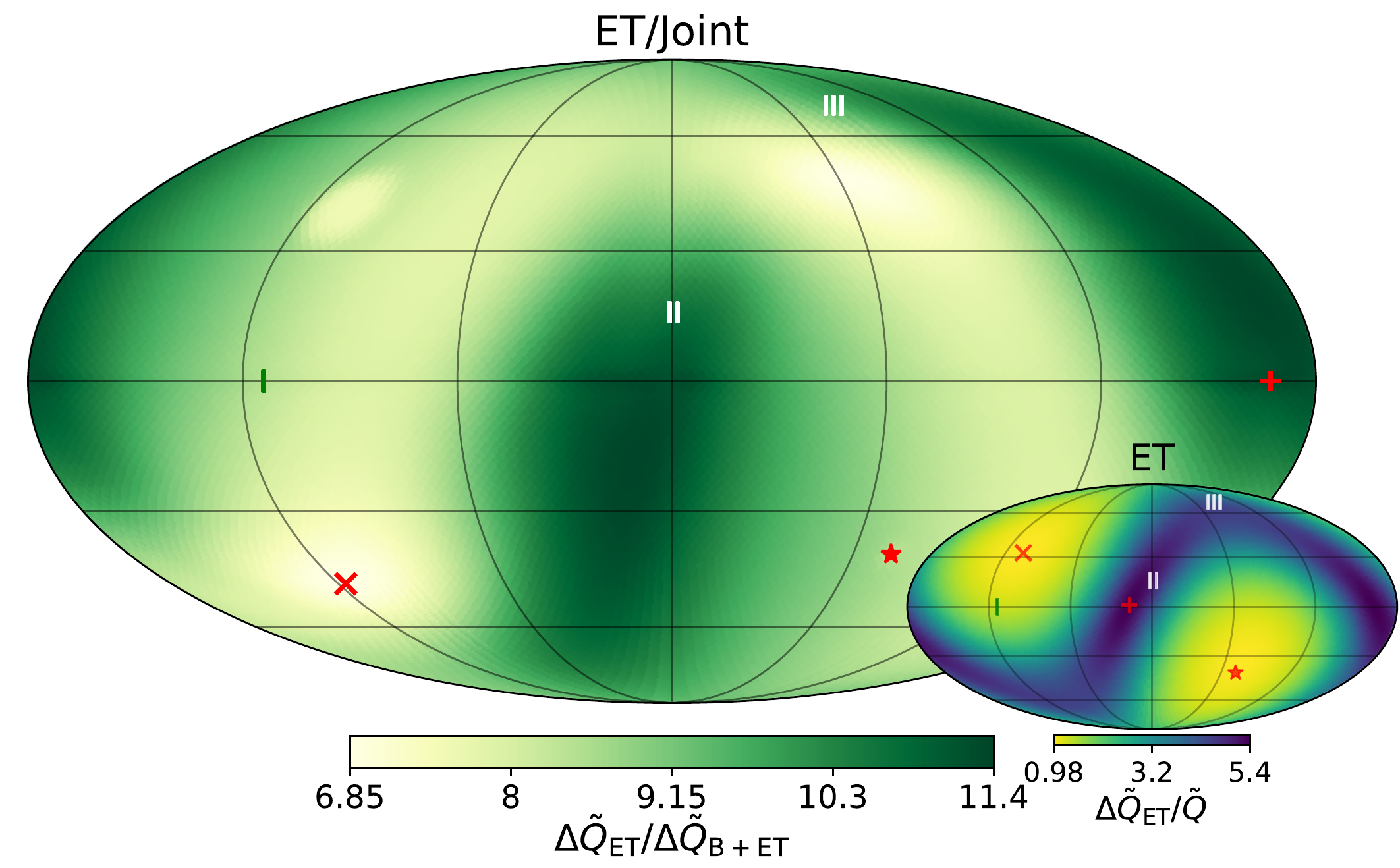}{0.5\textwidth}{(c)}}
\vspace{-0.4cm}
\caption{The combined constraints on the quadrupole parameter $\tilde Q$ for the
BNS system and its multiband improvement compared to a single detector case. See
the caption of Fig.~\ref{fig:SNR} for the meanings of marks, including
``$\star$'', ``$\times$'', ``$+$'', and Roman numbers.  (a) The combined
constraints on $\Delta\tilde Q/\tilde Q$ using ``B-DECIGO+ET''.  (b) Large: the
improvement of constraining $\Delta\tilde Q$ using ``B-DECIGO+ET'' to B-DECIGO
alone; Small: the constraints on $\Delta\tilde Q/\tilde Q$ using B-DECIGO alone.
(c) Large: the improvement of constraining $\Delta\tilde Q$ using ``B-DECIGO+ET'' to
ET alone; Small: the constraints on $\Delta\tilde Q/\tilde Q$ using ET alone.}
\label{fig:Q}
\end{figure}

\begin{figure*}%[h]
	\gridline{
	\fig{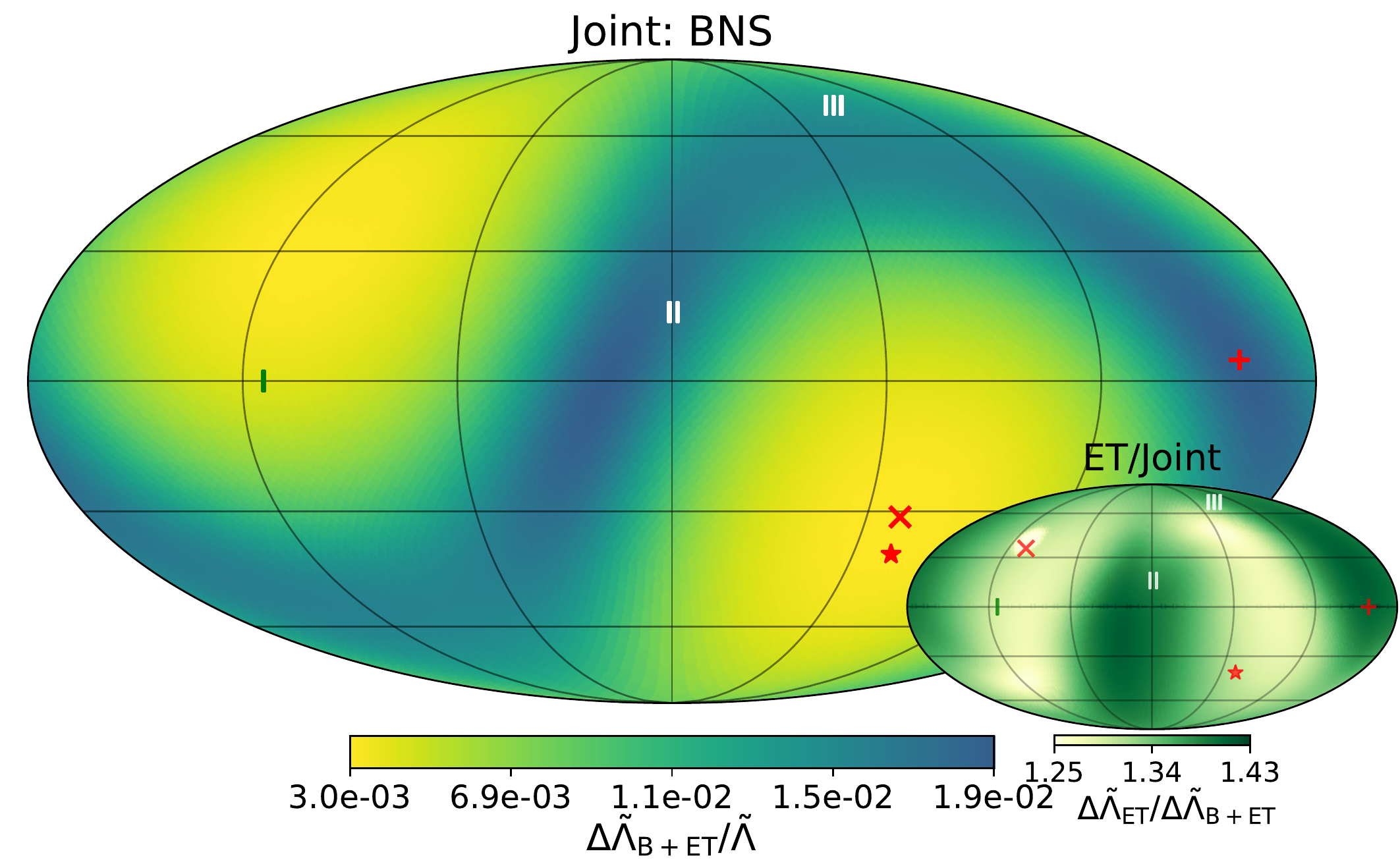}{0.5\textwidth}{(a)}
	,\fig{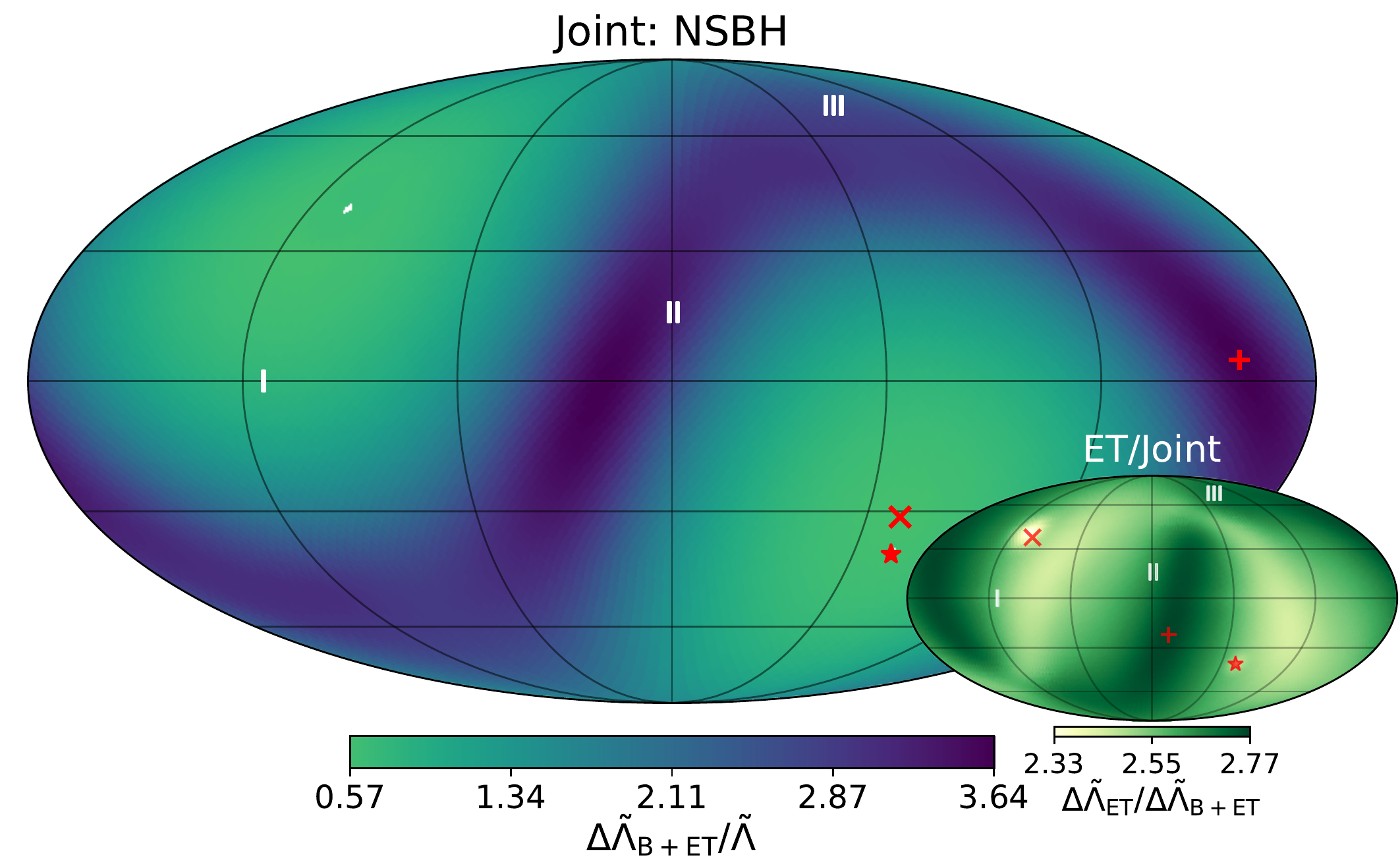}{0.5\textwidth}{(b)}}
	\caption{The combined constraints on tidal parameter $\tilde\Lambda$, and
	its multiband improvement compared to the limits from ET alone. See the
	caption of Fig.~\ref{fig:SNR} for the meanings of marks, including
	``$\star$'', ``$\times$'', ``$+$'', and Roman numbers.  (a) Large: the
	combined constraints on $\Delta\tilde \Lambda/\tilde \Lambda$ for the BNS
	system using ``B-DECIGO+ET''; Small: the improvement of constraining
	$\Delta\tilde \Lambda$ using ``B-DECIGO+ET'' to ET alone.  (b) Large: the
	combined constraints on $\Delta\tilde \Lambda/\tilde \Lambda$ for the NSBH
	system using ``B-DECIGO+ET''; Small: the improvement of constraining
	$\Delta\tilde \Lambda$ using ``B-DECIGO+ET'' to ET alone.}
	\label{fig:L}
\end{figure*}

We have also verified that all the parameters in $\bm{\Xi}^{\rm int}$ follow the
same distribution patterns on sky maps for each detector, similar to those of
$\Delta\tilde Q$ in Fig.~\ref{fig:Q} and $\Delta\tilde \Lambda$ in
Fig.~\ref{fig:L}.  Nevertheless, the precision distribution for extrinsic
parameters are different from one another.

In addition, we have executed a Fisher matrix analysis without estimating the
localization parameters ($\bar\phi_S,\bar\theta_S$), which can be seen as the PE
precision when the EM counterparts are observed and  they provide precise sky
location of the sources. We found in that case all errors for parameters in
$\bm{\Xi}^{\rm int}$ are about $80\%$ ($99\%$) of the errors in
Table~\ref{tab:result} for BNS II using B-DECIGO (ET), which is insignificant
compared to the improvement of multiband observations with the consideration of
localization.

%---------------------------------------------------------------------
%---------------------------------------------------------------------

%---------------------------------------------------------------------
\subsubsection{Quadrupole Parameter}
\label{subsec:Q}
%---------------------------------------------------------------------

We now focus on the quadrupole parameter, and show how joint detections benefit
the observations from space and ground in a mutual way.

In Fig.~\ref{fig:Q} (a) we show the combined constraints from ``B-DECIGO+ET''.
The relative error $\Delta \tilde Q_{\rm {B+ET}}/\tilde Q$ could reach down to
0.1 with the worst cases $< 0.8$, which indicates that $\tilde Q$  is measurable
no matter of the GW source direction. In contrast, the small sky maps in panels
(b) and (c) of Fig.~\ref{fig:Q} reveal that individual observations from neither
B-DECIGO nor ET can fully identify this parameter. Comparing the small map in
Fig.~\ref{fig:Q} (c) with Fig.~\ref{fig:Q} (a), we demonstrate that the sky
distribution of the combined precision, $\Delta \tilde Q_{\rm {B+ET}}/\tilde Q$,
as well as other intrinsic parameters as we have verified, is dominated by ET's
distribution pattern, $\Delta \tilde Q_{\rm {ET}}/\tilde Q$.

\begin{figure*}
\gridline{\fig{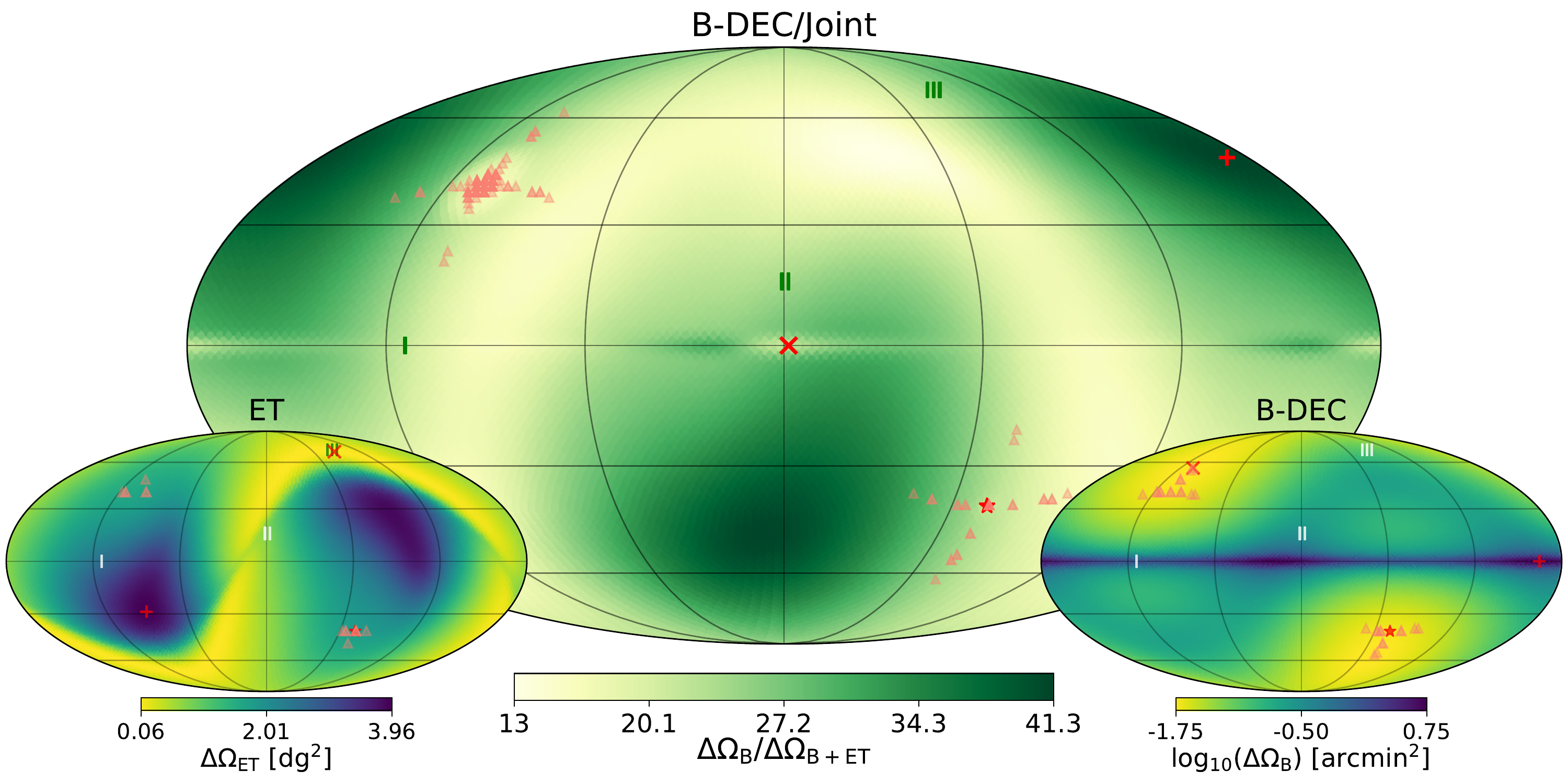}{0.7\textwidth}{}}
\caption{The localization ability of the BNS system.  Large: the localization
precision improvement using joint detection of ``B-DECIGO+ET'' than using
B-DECIGO alone.  Small: localization precision using ET (left) and B-DECIGO
(right) alone. The red triangles mark the location where parameter correlations
are larger than 0.9995. See the caption of Fig.~\ref{fig:SNR} for the meanings
of other marks, including ``$\star$'', ``$\times$'', ``$+$'', and Roman numbers.}
\label{fig:loc_B}
\end{figure*}

The large sky maps in panels  (b) and (c) of Fig.~\ref{fig:Q} stress the multiband
effects by plotting the {\it relative improvement} in comparison with using B-DECIGO
or ET alone. We notice that their distribution patterns are complementary. The
colorbar value shows that the joint detection measures 10--50 times better than
B-DECIGO alone and 6--12 times better than ET alone. We also notice that the
enhancement is less evident (the yellow region) usually at the location where
the former detector measures pretty well and the later joined detector measures
poorly.  There is a small region near the anti-aligned direction of
$\hat{\boldsymbol{L}}$ that has a relative lower improvement, which is caused by
a degeneracy of location parameters. Such degeneracy gets worse when joining
B-DECIGO with ET, therefore it leads to a worse PE improvement (see triangle
markers in Fig.~\ref{fig:loc_B}). 

%---------------------------------------------------------------------
%---------------------------------------------------------------------

Comparing the distribution of PE errors, e.g. in Fig.~\ref{fig:Q}~(a), with the
SNR in Fig.~\ref{fig:SNR}, we find the former has larger yellow areas, which also
indicates an imperfect inverse relation between PE errors and SNR. Even if some
locations have lower SNR, they still yield relative precise PE results.

From the measurement of quadrupole parameter, we see the striking advantage of
multiband detection, because $\Delta \tilde Q$ determined by a space-borne
detector or a ground-based detector alone is too large to yield any constraints,
while multiband observations enable pretty good limits on $\Delta \tilde Q$.

%---------------------------------------------------------------------
\subsubsection{Tidal Deformability}
\label{subsec:L}
%---------------------------------------------------------------------

We now investigate the multiband constraints on the tidal deformability, as
well as the comparison between BNS and NSBH systems. In Fig.~\ref{fig:L}, we
show the joint detection errors and multiband enhancement relative to using only
ET. 

Figure~\ref{fig:L} (a) displays the distribution of $\Delta\tilde\Lambda_{\rm
{B+ET}}/\tilde\Lambda$ for the BNS system as a function of sky location, which
gives a value ranging between $3\times 10^{-3}$ to $2\times 10^{-2}$.  The small
sky map indicates that multiband limits are dominated by ET's value, with 1--2
times tighter when B-DECIGO joins in. In contrast to other intrinsic parameters,
the tidal effect starts from 5\,PN and contributes largely at the very last
stage of inspiral in the ET band, therefore ET plays a leading role in
constraining $\tilde\Lambda$.

Figure~\ref{fig:L} (b) is the NSBH analog to Fig.~\ref{fig:L} (a). Comparing the
large maps in both panels, we see that the BNS system yields a tighter relative
error, $\Delta\tilde\Lambda_{\rm {B+ET}}/\tilde\Lambda$, while NSBH system can
barely measure the tidal deformability, even in the multiband case. To
compare them, we select BNS II and NSBH II, and normalize the precision to the
same distance $D_L$,
%--
\begin{align}
	&\Big[\frac{\Delta\tilde\Lambda_{\rm {B+ET}}}{\tilde\Lambda}\Big]_{\rm BNS} = 0.017\Big(\frac{D_L}{40\,{\rm Mpc}}\Big) \,,
	\\
	&\Big[\frac{\Delta\tilde\Lambda_{\rm {B+ET}}}{\tilde\Lambda}\Big]_{\rm NSBH} = 0.47\Big(\frac{D_L}{40\,{\rm Mpc}}\Big) \,.
\end{align}
% \begin{align}
% 	&\Big[\frac{\Delta\tilde\Lambda_{\rm {B+ET}}}{\tilde\Lambda}\Big]_{\rm BNS} = 0.017\Big(\frac{D_L}{40\,{\rm Mpc}}\Big) = 0.12\Big(\frac{D_L}{280\,{\rm Mpc}}\Big)\,,
% 	\\
% 	&\Big[\frac{\Delta\tilde\Lambda_{\rm {B+ET}}}{\tilde\Lambda}\Big]_{\rm NSBH} = 0.47\Big(\frac{D_L}{40\,{\rm Mpc}}\Big) = 3.3\Big(\frac{D_L}{280\,{\rm Mpc}}\Big)\,.
% \end{align}
%--
Such difference partially comes from the fact that BNS has a larger tidal
deformability parameter $\tilde{\Lambda}$, which leads to a larger contribution
of the tidal term in the GW phase. The detectors are thus more sensitive to
$\tilde\Lambda$. Another reason is that BNS signal has a longer time (5.6\,d)
in ET than NSBH (0.9\,d), which helps it accumulate more GW cycles, hence ET
could extract more  information of the tidal parameter. 

The improvement ratio $\Delta\tilde\Lambda_{\rm {ET}}/\Delta\tilde\Lambda_{\rm
{B+ET}}$, on the other hand, is more significant in NSBH system with more
information gained from B-DECIGO.  From the NSBH II we notice that even when
B-DECIGO measures $\Delta\tilde\Lambda_{\rm {B}}/\tilde\Lambda$ larger than
$10^3$, in multiband it still helps reduce ET's uncertainty by about 3 times.
Such improvement benefits from the precise measurements of B-DECIGO on other
parameters, which helps ET to break the  degeneracies among them. It is similar
to the measurement of the dipole radiation parameter in~\citet{Zhao:2021bjw}.

The above analysis of comparing BNS and NSBH systems could extend to other
parameters as well. For the length of discussion, we do not give too much detail
here.

%---------------------------------------------------------------------
%---------------------------------------------------------------------

%---------------------------------------------------------------------
\subsection{Estimation on Extrinsic Parameters}
\label{sec:ext}
%---------------------------------------------------------------------

In this subsection, we demonstrate how multiband observations improve extrinsic
parameters in $\bm{\Xi}^{\rm ext}$, especially focusing on the angular
resolution $\Delta \Omega$. Same as the last subsection, we begin with some
general characteristics of $\bm{\Xi}^{\rm ext}$ and then concentrate
respectively in Sec.~\ref{sec:Omega} and Sec.~\ref{sec:alert} on $\Delta
\Omega$, and its improvement over time which is important for EM follow-ups.

Worth to note that, in contrast to the intrinsic parameters in $\bm{\Xi}^{\rm
int}$, the sky distribution pattern of each parameter in $\bm{\Xi}^{\rm ext}$ is
not the same, especially among parameters $D_L$, $t_c$, $\bar\phi_S$, and
$\bar\theta_S$, though we only show the distribution of $\Delta \Omega$ in this
study. 

The time at coalescence $t_c$ is an intriguing parameter that correlates with
both intrinsic and extrinsic parameters, as can be seen in the correlation
matrices in Fig.~\ref{fig:corr}. Therefore measurement of $t_c$ will be affected
from many aspects. For example, if we know the location of the source from EM
observations---which means no need to estimate $\bar\phi_S$ and
$\bar\theta_S$---unlike $\bm{\Xi}^{\rm int}$, the precision of $t_c$ will
improve enormously by nearly an order of magnitude. Moreover, although whether
or not to include the Earth's orbit in the Doppler phase $\varphi_{D}(t)$ does
not affect the PE results of all other parameters, $\Delta t_c$ will be
influenced by about one order of magnitude.

 Also unlike the nearly inverse relation between intrinsic parameters
 $\bm{\Xi}^{\rm int}$ and the SNR, the value $\Delta \Omega$ is largely
 unrelated with SNR. From column (12) in Table~\ref{tab:result} we see that
 among three locations, location I has the largest localization area, but
 highest SNR.  Comparing B-DECIGO's PE results of BNS II and BNS III, they have
 similar SNRs, $\Delta\bm{\Xi}^{\rm int}$, $\Delta t_c$, and $\Delta\phi_c$, but
 significant differences in $\Delta D_L$ and $\Delta \Omega$. The precision
 $\Delta D_L$ is co-determined by the SNR and the sky location of the source.

%---------------------------------------------------------------------
\subsubsection{Sky Localization}
\label{sec:Omega}
%---------------------------------------------------------------------

Due to the need of an accurate sky location for successful EM follow-ups, we pay
close attention to the precision of $\Delta \Omega$. We present the multiband
sky localization improvement as well as constraints from individual detectors
for the BNS system in Fig.~\ref{fig:loc_B}. The reason why we display the
improvement $\Delta\Omega_{\rm B}/\Delta\Omega_{\rm B+ET}$ other than
$\Delta\Omega_{\rm B+ET}$ is that the distribution of $\Delta\Omega_{\rm B+ET}$
is completely dominated by $\Delta\Omega_{\rm B}$'s pattern, which means that
the uncertainty distribution of $\Delta\Omega_{\rm B+ET}$ is extremely close to
the right small map in Fig.~\ref{fig:loc_B}. Note that in these sky maps of
$\Delta\Omega$, we smooth the values on the map over a few pixels to eliminate
the discrete effect caused by a limited number of points. The smoothing leads to
a decrease of the maximum value in the B-DECIGO plot, but does not affect the
discussions that we have here.

Comparing the two small maps in Fig.~\ref{fig:loc_B}, we notice that B-DECIGO
can localize down to $10^{-2}$\,arcmin$^2$, which is orders of magnitude better
than ET. ET, albeit less powerful, still exceeds most ground-based observatories
for it has three individual detectors and a lower cut-off frequency which
prolong the signal's active time in the sensitive band. The localization
precisions, as well as sky distributions, are mainly determined by the motions
and orbital baselines of the two detectors.  The multiband improvement
$\Delta\Omega_{\rm B}/\Delta\Omega_{\rm B+ET}$ ranges from $10$ to $50$, which
is far better than that of $\Delta\tilde \Lambda_{\rm ET}/\Delta\tilde \Lambda_{\rm
B+ET}$. Considering that both $\Delta\Omega_{\rm ET}$ and $\Delta\tilde
\Lambda_{\rm B}$ are 3--4 orders of magnitude worse than $\Delta\Omega_{\rm B}$
and $\Delta\tilde \Lambda_{\rm ET}$, such a huge improvement in sky localization
benefits from the two distinct orbits of B-DECIGO and ET.

%---------------------------------------------------------------------
\begin{figure}
	\centering
	\includegraphics[width=1.0\linewidth]{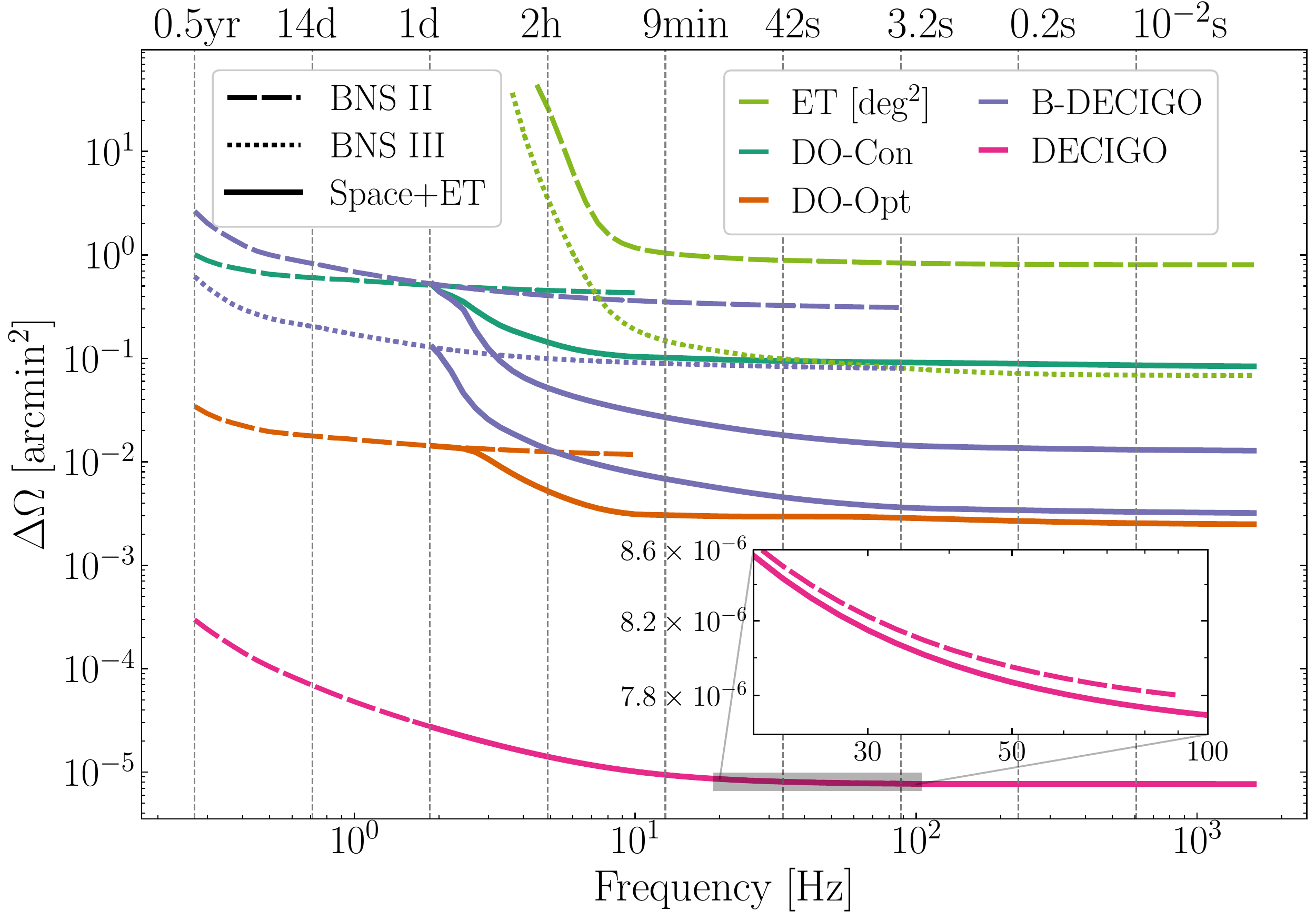}
\caption{Localization precision $\Delta\Omega$ as a function of frequency (time)
for the BNS II and BNS III systems. The dashed (location II) and dotted
(location III) lines are the results for single detectors and the solid lines
are the results for the joint detection of the corresponding space-borne
detectors and ET, which starts from 2\,Hz. Note that for ET alone (light green
lines), $\Delta\Omega$ is in the unit of square degrees.}
\label{fig:loc_time}
\end{figure}
%---------------------------------------------------------------------

\begin{figure*}%[h]
	\gridline{
	\fig{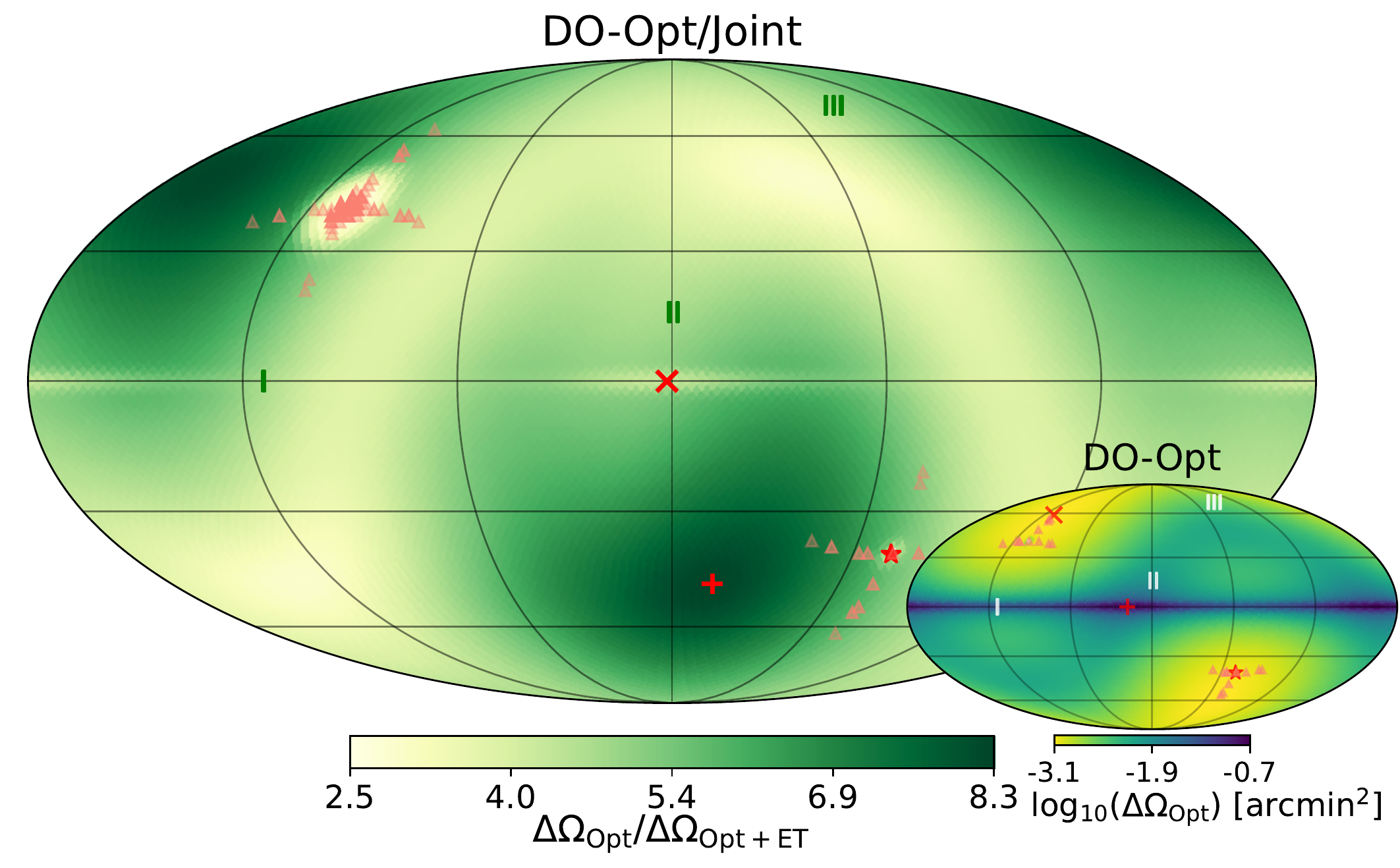}{0.5\textwidth}{(a)}
	\fig{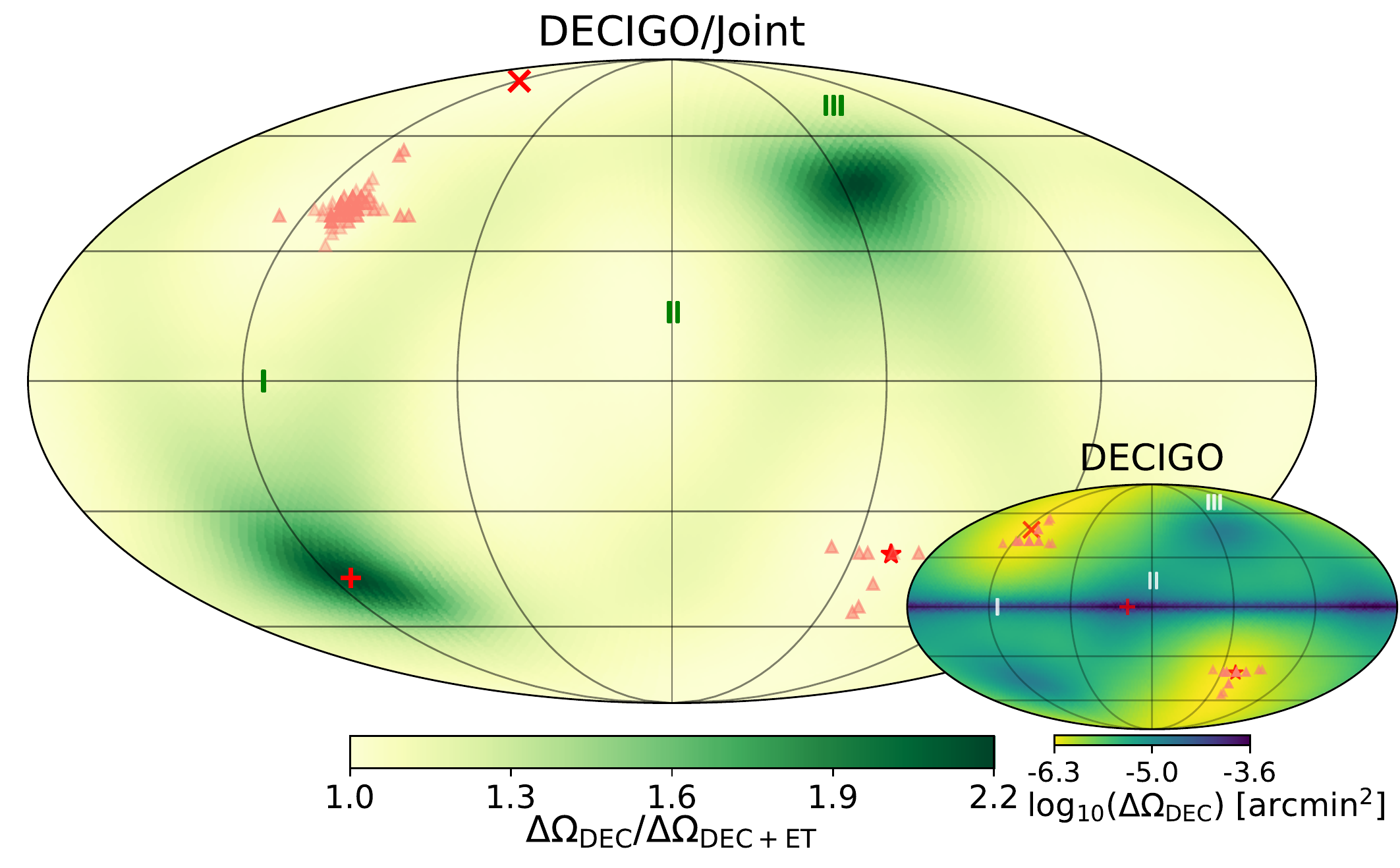}{0.5\textwidth}{(b)}
	}
	\vspace{-0.4cm}
	\gridline{
	\fig{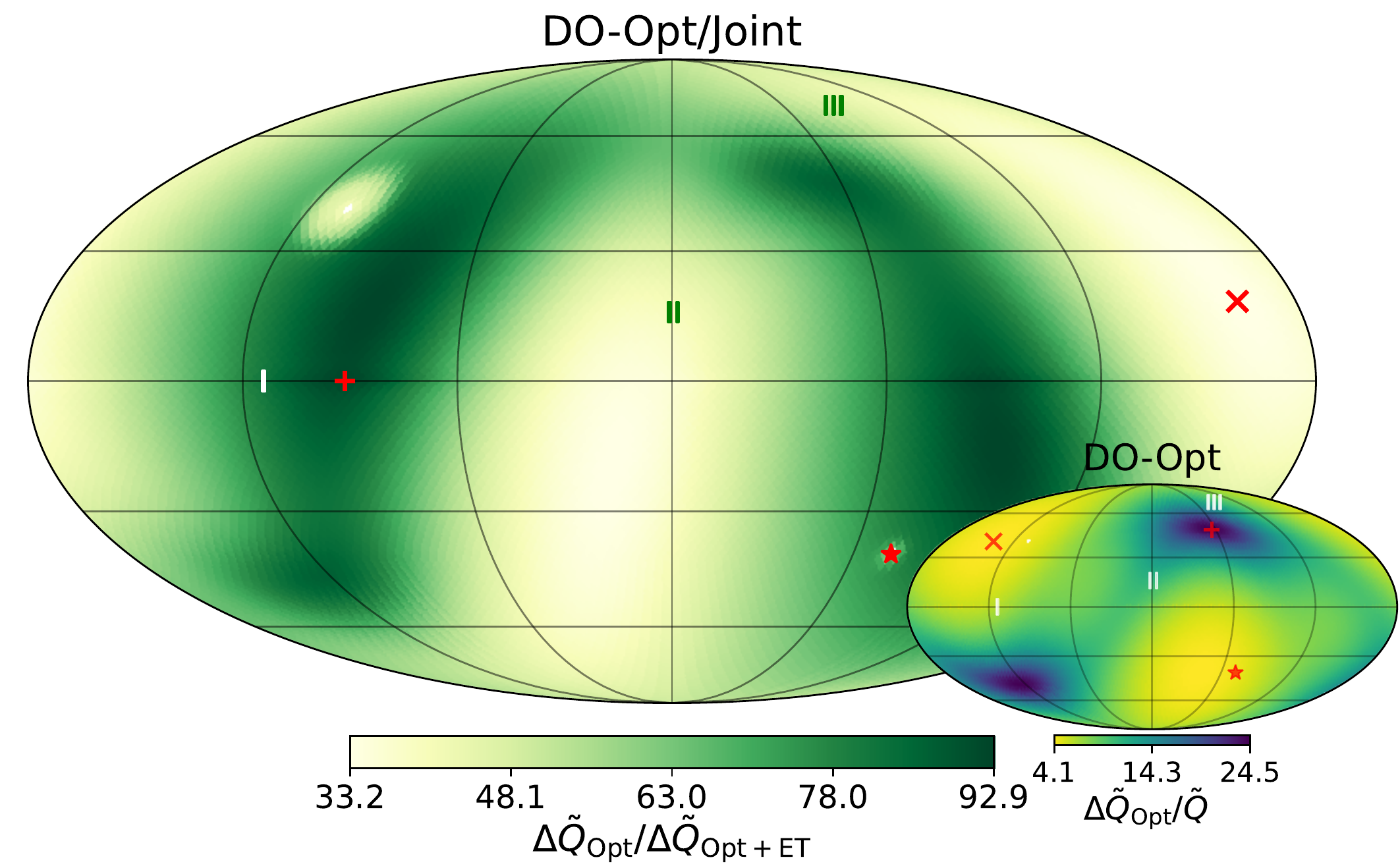}{0.5\textwidth}{(c)}
	\fig{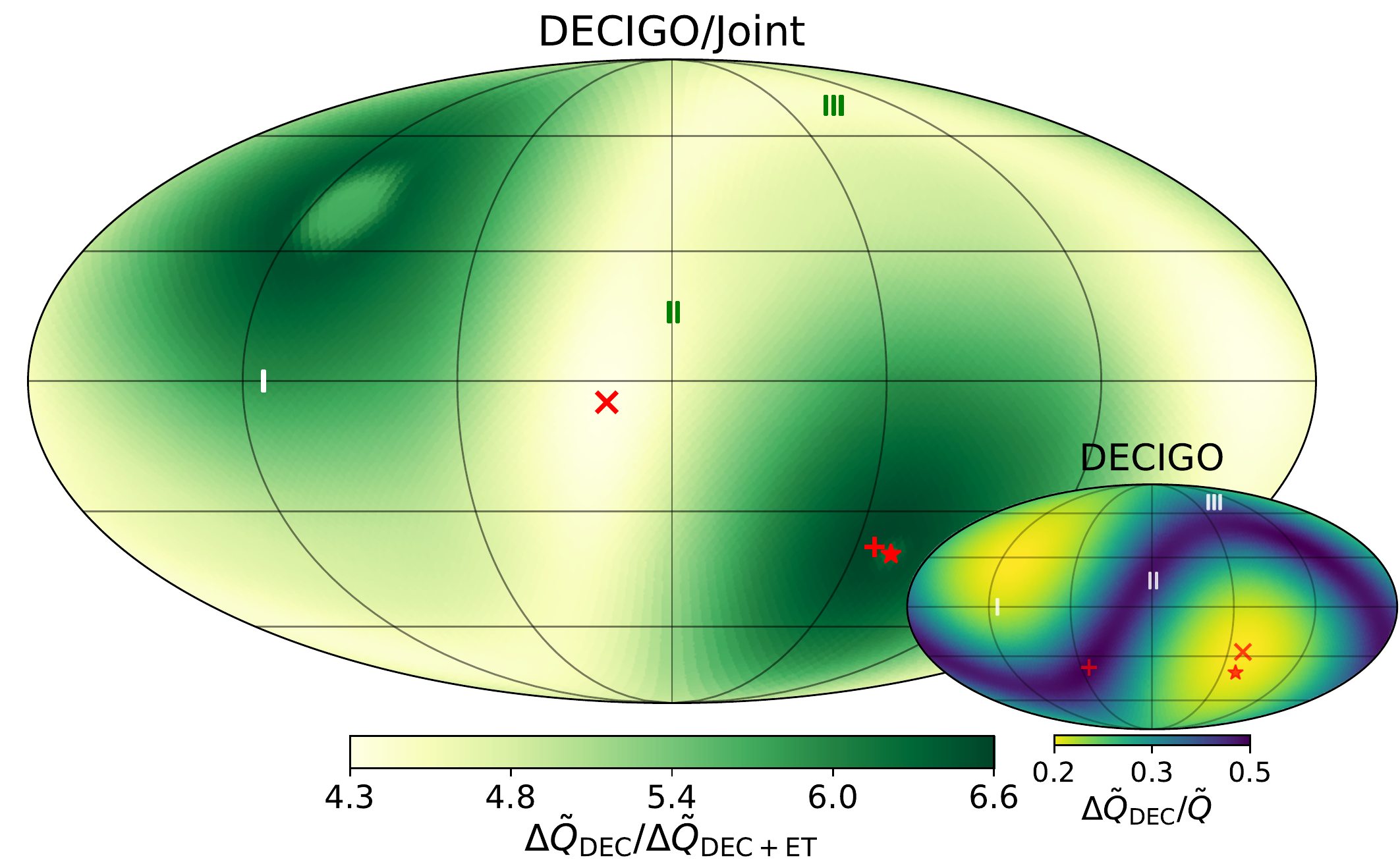}{0.5\textwidth}{(d)}
	}
	\vspace{-0.4cm}
	\caption{The multiband improvement and constraining results of
	$\Delta\Omega$ [(a) and (b)] and $\Delta\tilde Q/\tilde Q$ [(c) and (d)] for
	the BNS system using DO-Optimal [(a) and (c)] and DECIGO [(b) and (d)].  The
	red triangles mark the location where parameter correlations are larger than
	0.9995. See the caption of Fig.~\ref{fig:SNR} for the meanings of other marks,
	including ``$\star$'', ``$\times$'', ``$+$'', and Roman numbers.  (a/b)
	Large: the improvement of constraining $\Delta\Omega$ using
	DO-Optimal/DECIGO together with  ET  to  the use of DO-Optimal/DECIGO alone;
	Small: the constraints on $\Delta\Omega$ using DO-Optimal/DECIGO alone.
	(c/d) Same as (a/b), for the quadrupole parameter $\tilde{Q}$.}
	\label{fig:other_det}
\end{figure*}

From the B-DECIGO plot in the lower right of  Fig.~\ref{fig:loc_B}, we notice
large errors in the ecliptic plane. This is a unique effect for the space-borne
detectors as it also shows up in Fig.~\ref{fig:other_det} for DOs and DECIGO.
Such a  phenomenon results from a combination of detector orbits and the signal
duration.  We first clarify that $\Delta\Omega$ is co-determined by
$\Delta\bar\mu_S$ and $\Delta\bar\phi_S$, and this ``line-like'' effect comes
from the characteristic of $\bar\mu_S$.

When the ecliptic polar angle of the source $\bar\theta_S$  changes from
$0^\circ/180^\circ$ (the two poles) to $90^\circ$ (the ecliptic plane), the
partial derivative of ${\partial \tilde{h}}/{\partial\bar\mu_S}$ turns smaller
across three orders of magnitude.  With the weight $S_n$ in
Eq.~\eqref{eq:innerproduct}, the integral $\Gamma_{\bar\mu_S\bar\mu_S}$ becomes
smaller by eight orders of magnitude, which makes the error $\Delta\bar\mu_S$
ranges from $10^{-5}$ to $10^{-2}$ gradually from the two poles to the ecliptic
plane. While $\Delta\bar\phi_S$ stays around $10^{-5}$, the angular resolution
is determined by the less accurate $\Delta\bar\mu_S$; accordingly, large
localization errors appear near the ecliptic plane.

However, such ``line-like'' effect will eventually vanish with the increase of
the source masses, thanks to another characteristic from $\bar\phi_S$.  With the
increment of mass, GW source merges at a lower frequency, which makes the GW
signal exists in the detector's band for a  shorter time---the signal enters the
detectors only few months or days before coalescence for decihertz detectors.
This is a key factor because ${\partial \tilde{h}}/{\partial\bar\phi_S}$, which
depends strongly on the secular evolution of the detector orbit, is orders of
magnitude lower after one month before merger. Without the information of the
long-term orbital modulation, the integral $\Gamma_{\bar\phi_S\bar\phi_S}$ is
tremendously smaller, which leads to a higher $\Delta\bar\phi_S$. While
$\Delta\bar\mu_S$ still varies greatly with latitude, it no longer dominates the
value of $\Delta\Omega$. Therefore the ``line-like'' effect vanishes.

We have also verified that for sources whose total mass $M \gtrsim 10^3
M_\odot$, this bad measurement along the ecliptic plane disappears. From a
mathematic point of view, the whole process is a competition between ${\partial
\tilde{h}}/{\partial\bar\mu_S}$, ${\partial \tilde{h}}/{\partial\bar\phi_S}$, and
$S_n$.

We mark the locations, where the maximum correlation in the off-diagonal
elements of the correction matrix is over 0.9995, in each map in
Fig.~\ref{fig:loc_B} by red triangles. We illustrate them briefly in
Sec.~\ref{sec:corr} and conclude that ET is slightly better than B-DECIGO in
breaking such degeneracy and joint observations deepen this effect by enlarging
the regions covered by red triangles.

%---------------------------------------------------------------------
\subsubsection{Early Warnings}
\label{sec:alert}
%---------------------------------------------------------------------

%---------------------------------------------------------------------
%---------------------------------------------------------------------

Prompt communication of source location from the GW detection to EM facilities
is crucial for multi-messenger follow-ups. While the accurate sky area derived
from the complete GW signal is promising, the pre-merger alert is equally
important. 

We show the localization precision of the BNS systems as a function of frequency
in Fig.~\ref{fig:loc_time}, and mark the time before coalescence at the top of
the plot. We find that with the joining in of ET, the localization precision
$\Delta\Omega$ gradually narrows. The shape of how $\Delta\Omega$ improves
depends on the detectors' sensitivity curves and designs. In addition, the same
shape of the dashed and the dotted blue lines reveals that the location of the
source do not affect the variation trend of the multiband improvement.

With nearly 4 years' time, the space-borne detector already has a stable
localization area $\sim \rm{arcmin^2}$. ET starts the observation about 5 days
before the merger, and it will gradually narrow down its own localization area
until a few minutes before merger to be stable, as shown in the light green
lines.  However, after ET joins the observation, it will not improve the sky
area of the space-borne detectors in the first few days. At one day before the
merger, the joint detection begins to take effect;  $\Delta\Omega$ gradually
drops again and finally has an improvement of 1--2 orders of magnitude for DOs
and B-DECIGO. 

DO-Conservative and B-DECIGO have similar single detector angular resolutions,
but different multiband improvements. ``B-DECIGO+ET'' is about one order of
magnitude better than ``DO-Conservative+ET'', which is caused by the better
sensitivity of B-DECIGO in the high frequency band. Furthermore, from the blue
and dark green dashed lines in Fig.~\ref{fig:loc_time} we discover that the
variation trend of $\Delta\Omega$ with frequency is different: B-DECIGO drops
more sharply, despite B-DECIGO and DOs have the same orbital configuration.  The
reason might be that, at a frequency larger than 0.1\,Hz, the BNS signal is at the
lowest noise region, the so-called sweet point, of B-DECIGO's sensitivity curve.
Therefore it gains more information from the source after this frequency while
DOs get more information before this point. 

The multiband enhancement of DECIGO, on the other hand, is different from the
other three decihertz detectors. Benefiting from four LISA-like designs, itself
can localize precisely to $\sim 10^{-6}~\rm{arcmin^2}$. The sky area keeps
shrinking until minutes before merger and the inclusion of ET barely improves
the precision. We will discuss more about comparisons between different
decihertz detectors in the next subsection.

%---------------------------------------------------------------------
\subsection{Comparison Between Different Decihertz Detectors}
\label{sec:other_det}
%---------------------------------------------------------------------

In this subsection, we compare the multiband results by combining ET with
different decihertz observatories.  In Fig.~\ref{fig:other_det}, we plot the
DO-Optimal and DECIGO analogs to Fig.~\ref{fig:Q} (b) and Fig.~\ref{fig:loc_B}
in order to highlight their similarities and differences. 

For the sky localization ability, from the small maps in
Fig.~\ref{fig:other_det} (a), Fig.~\ref{fig:other_det} (b), and
Fig.~\ref{fig:loc_B}, we see that the sky distribution is similar, except DECIGO
appears more clumpy with more apparent delimitations.  DOs have similar designs
with B-DECIGO, thus the localization precision is approximately $\propto 1/{\rm
SNR}^2$. Whereas DECIGO has exceedingly good localization precision down to
$10^{-6}\,{\rm arcmin^2}$ because of multiple interferometers. This distinction
has also been reflected in the large maps in Fig.~\ref{fig:other_det} (a) and
Fig.~\ref{fig:other_det} (b), where the difference of multiband enhancement is
displayed. From localization point of view, DECIGO's ability is sufficient by
itself.

For intrinsic parameters, based on the measurement on $\Delta\tilde Q/\tilde Q$,
we stress three points.  (i) Comparing the small map in Fig.~\ref{fig:other_det}
(c) with the small map in Fig.~\ref{fig:Q} (b), despite DO-Optimal yields larger
SNRs, its measurements on $\tilde Q$ do not exceed that of B-DECIGO, because DOs
have a relatively poor performance at high frequencies. This situation also
applies for $\tilde \Lambda$ that enters at higher PN order.  (ii) Comparing the
small map in Fig.~\ref{fig:other_det} (d) with the small map in Fig.~\ref{fig:Q}
(b), we notice that with the help of four interferometers, DECIGO alone could
discriminate the quadrupole parameter. Moreover, the uncertainty dispersion on
the sky map reduces significantly.  (iii) Comparing the large map in
Fig.~\ref{fig:other_det} (d) with the small map in Fig.~\ref{fig:Q} (c), we see
that the relative improvement $\Delta\tilde Q_{\rm DEC}/\Delta\tilde Q_{\rm
DEC+ET}$ just follows the distribution of $\Delta\tilde Q_{\rm ET}$, which
indicates that DECIGO has reached its ceiling in detecting $\Delta\tilde Q$.

%---------------------------------------------------------------------
\begin{figure}
\gridline{
\fig{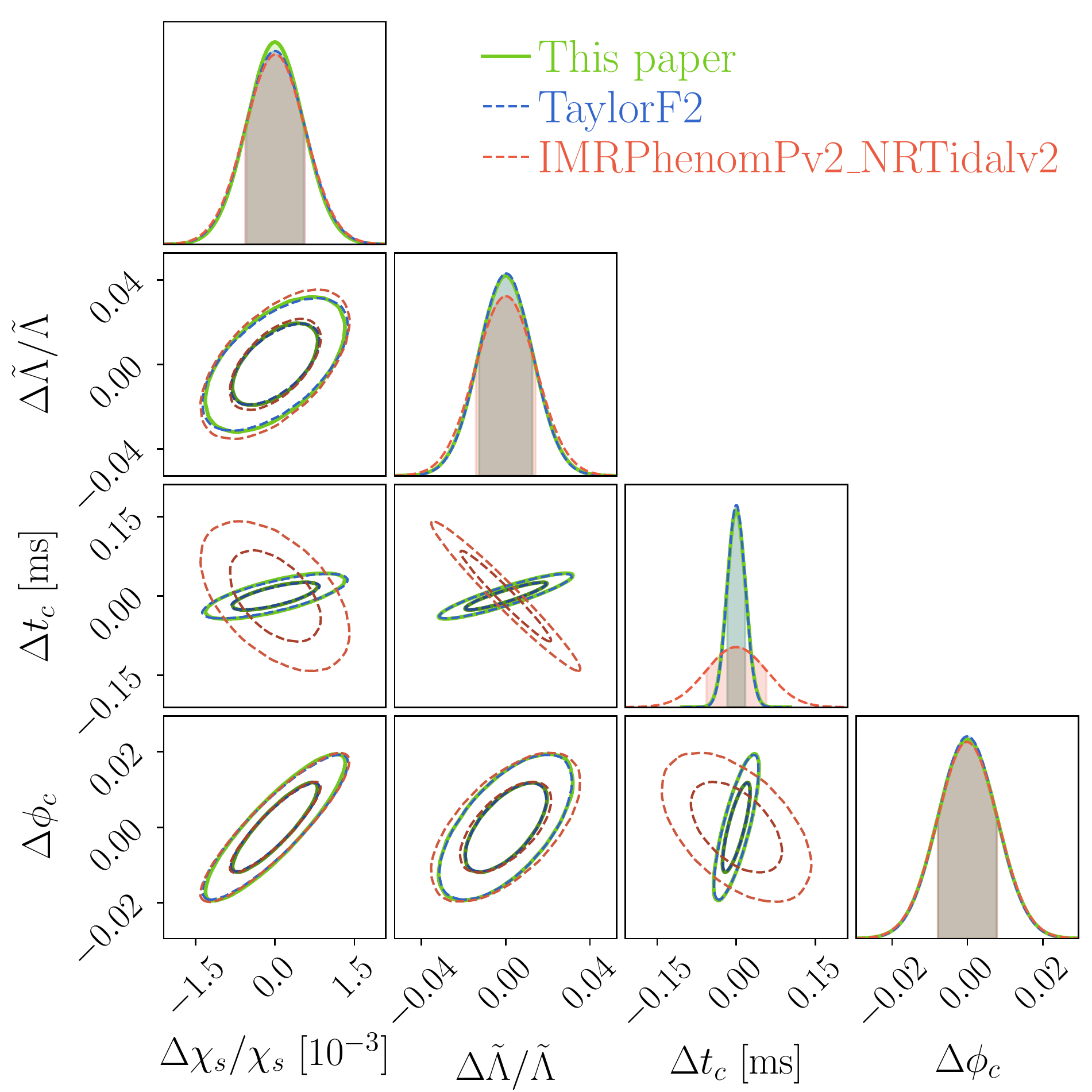}{0.45\textwidth}{}}
\caption{The $1\,\sigma$ and $2\,\sigma$ constraints from ``B-DECIGO+ET'' for BNS II using
different waveform models.}
\label{fig:wf_compare}
\end{figure}
%---------------------------------------------------------------------

%---------------------------------------------------------------------
\subsection{Comparison Between Different Waveforms}
\label{sec:PV2}
%---------------------------------------------------------------------

After investigating the PE results constrained by different detectors, we extend
our method to other waveform models and confirm that the PN waveform and
phenomenological waveform could yield similar statistical errors.

We adopted two templates implemented in the LIGO Scientific Collaboration (LSC)
Algorithm Library \citep[LAL;][]{lalsuite}, \texttt{TaylorF2} and
\texttt{IMRPhenomPv2\char`_NRTidalv2} \citep{Husa:2015iqa,Dietrich:2019kaq}.
\texttt{TaylorF2} is the same with ours except including a tidal correction up
to 7\,PN.  \texttt{IMRPhenomPv2\char`_NRTidalv2} uses the precessing
phenomenological BBH waveform baseline, augmented with a tidal prescription
``NRTidalv2''. Other than PN approximation, ``NRTidalv2'' uses a
numerical-relativity-based closed-form tidal phase contribution with a tidal
amplitude correction, as well as an inclusion of spin-spin and cubic-in-spin
effects up to 3.5\,PN \citep{Dietrich:2017aum,
Dietrich:2018uni,Dietrich:2019kaq}. Furthermore, it implements the universal
relations to relate the tidal deformability to the spin-induced
quadrupole-monopole terms, and therefore reduces the parameters $\kappa_{1,2}$.
To compare our results with it, we change the parameter set to
%--
\begin{equation} 
\label{eq:PE2}
	\bm{\Xi}^{\rm PE} \equiv  \big\{  {\cal M}, \eta,	 \chi_s,\tilde\Lambda \big\} \cup \bm{\Xi}^{\rm ext} \,,
\end{equation}
%--
and inject only the spin-aligned waveforms in \texttt{IMRPhenomPv2\char`_NRTidalv2}.

In Fig.~\ref{fig:wf_compare}, we briefly compare the multiband parameter errors
of BNS II detected by ``B-DECIGO+ET'' adopting ``our TaylorF2'' in
Sec.~\ref{sec:wf}, \texttt{TaylorF2} in LAL, and
\texttt{IMRPhenomPv2\char`_NRTidalv2}.  We exhibit the results of four
parameters $\chi_s$, $\tilde \Lambda$, $\Delta t_c$, and $\phi_c$ that vary
among the three waveforms. For single-detector detections, such variation is
more evident in B-DECIGO and less obvious in ET, and the variation of the
combined errors we present here is a mixture of B-DECIGO's and ET's trend.  We
have checked that, the parameters that do not appear in the figure are
consistent across all three waveform models.  We also find that, without
assessing $\tilde Q$, $\Delta \Lambda$ has been tightened marginally by 30\%
than before, from $1.7\%$ to $1.3\%$. 

We notice that our lower-order TaylorF2 and \texttt{TaylorF2} in LAL are very
close to each other in the uncertainties and correlations, indicating a weak
impact of higher order corrections in this study.  In contrast,
\texttt{IMRPhenomPv2\char`_NRTidalv2} gives a wider bound, especially in $\Delta
t_c$, which may be induced by the complex correlations of $t_c$ with other
parameters and it also reminds us to be cautious when analyzing $\Delta t_c$.
The variation in $\Delta\tilde\Lambda$ demonstrates that waveform selection has
a small impact on statistical errors for the scenarios considered in this paper. 

Note that our purpose here is to tentatively compare the statistical errors
estimated by the Fisher analysis, which is over-simplified. For detailed
analyses on waveform comparison, see \citet{Narikawa:2019xng},
\citet{Gamba:2020wgg}, and \citet{Wade:2014vqa}.

%---------------------------------------------------------------------
\section{Constraining NS's EoS}
\label{sec:EoS}
%---------------------------------------------------------------------

In the above discussion, in particular in Sec.~\ref{subsec:L}, we have analyzed
the projected precision of measuring $\tilde \Lambda$. Ultimately, we are
interested in limiting individual tidal deformability $\Lambda_i$, which relates
more closely to the NS's EoS. Many works have paid attention to it. They have
pushed forward the limitation on EoS gradually with more accurate waveforms
\citep{Hinderer:2009ca, Read:2009yp, Damour:2012yf},  improved Bayesian method
\citep{DelPozzo:2013ala, Agathos:2015uaa},  consideration of a population of
BNSs \citep{Lackey:2014fwa}, and  updated detectors \citep{Yagi:2013awa,
Wang:2020xwn}. Using the 3G detectors, $\Delta\Lambda_i$ can be limited to an
accuracy better than 10\%.

%---------------------------------------------------------------------
\begin{figure}%[h]
	\includegraphics[width=\linewidth]{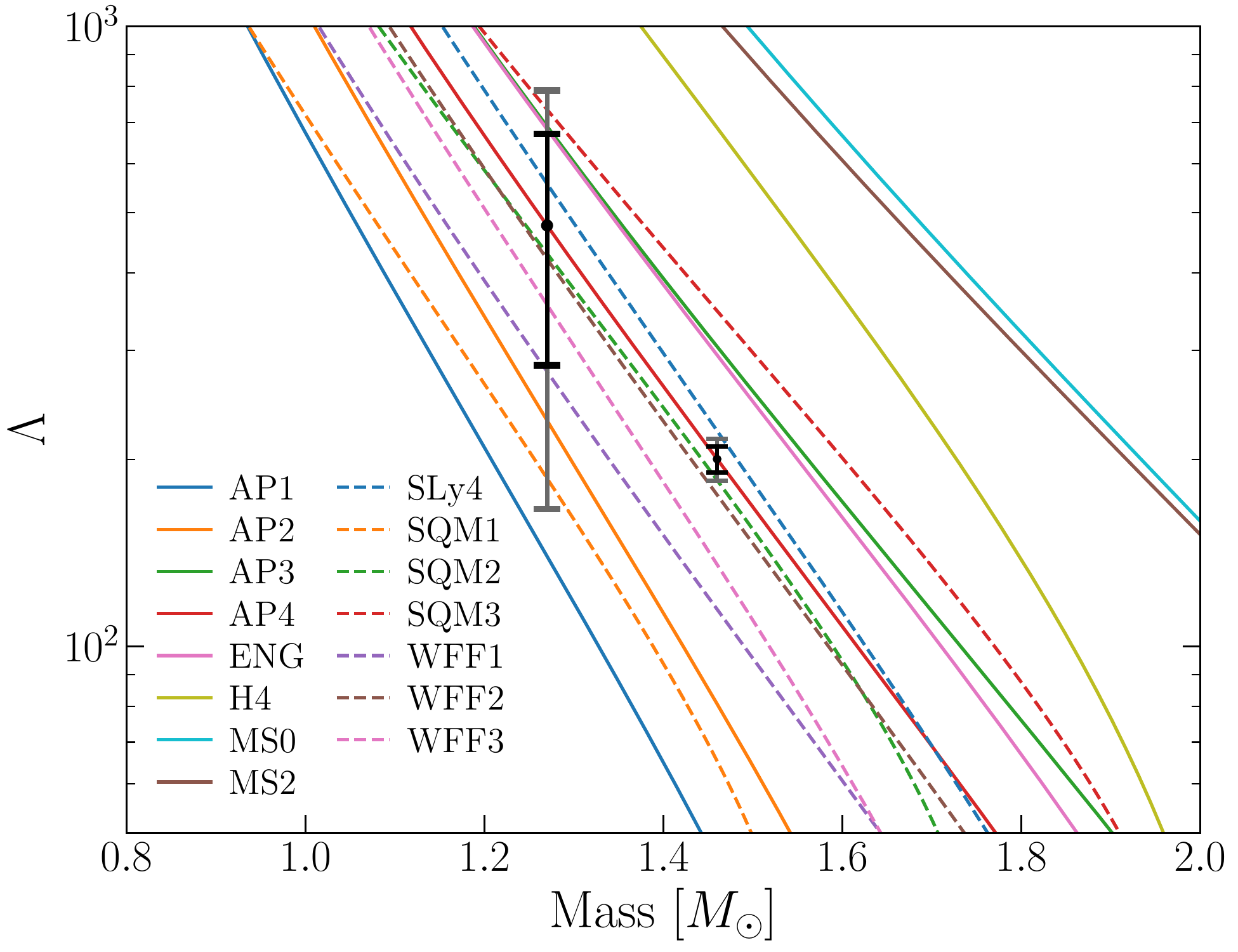}
	\caption{The projected 1\,$\sigma$ constraint at $m^{\rm S}_1 = 1.46
	M_{\odot}$, and the projected 10\,$\sigma$ constraint at $m^{\rm S}_2 = 1.27
	M_{\odot}$,  on the individual tidal deformability parameters
	$\Lambda_{1,2}$ from joint detection using ``B-DECIGO+ET'' (the black
	errorbars) versus using ET alone (the grey errorbars). Colored curves give
	the prediction of $\Lambda$ from different EoSs.}
	\label{fig:EoS}
\end{figure}
%---------------------------------------------------------------------

Following early works, we explore the the multiband constraints on $\Lambda_i$
here. By Taylor series expanding $\lambda_i$ about a fiducial mass $m_0^{\rm S} = 1.4
M_{\odot}$ \citep{Agathos:2015uaa}, one gets
%--
\begin{equation}
\lambda_i(m) \simeq c_{0}+c_{1}\left(\frac{m^{\rm S}_i-m^{\rm S}_{0}}{M_{\odot}}\right)+\frac{1}{2} c_{2}\left(\frac{m_i^{\rm S}-m^{\rm S}_{0}}{M_{\odot}}\right)^{2}\,.
\label{eq:lambda}
\end{equation}
%--
We could estimate the precision of $c_0$, and then obtain the limitation on
$\Lambda_i$. We take the EoS AP4 as an example to illustrate the multiband
constraints. The fiducial values of $[c_0,c_1,c_2]$ of AP4 are respectively
$[4.09, -3.87, -2.8]\times10^{-24}\,$s$^5$. By assuming that both NSs are described
by the same EoS, we use AP4's tidal deformability as the fiducial values for the
BNS system at location II and derive $\Delta\tilde\Lambda_{\rm
{B+ET}}/\tilde\Lambda=4.3\%$ and $\Delta\tilde\Lambda_{\rm
{ET}}/\tilde\Lambda=7.0\%$. Using Eq.~\eqref{eq:lambda}, we give our multiband
constraints on the uncertainty of $\Lambda_{1,2}$,
%--
\begin{align}
\Delta \Lambda_1 &= \rep{19}\left(\frac{D_L}{40\,{\rm Mpc}}\right) \,, \\
\quad \Delta \Lambda_2 &= \rep{10}\left(\frac{D_L}{40\,{\rm Mpc}}\right)\,.
\end{align}

In Fig.~\ref{fig:EoS} we show the multiband constraints on $\Lambda_{1,2}$. At
mass $m^{\rm S}_1 = 1.46 \, M_{\odot}$, we plot the 1\,$\sigma$ constraints by
``B-DECIGO+ET'' in black and by ET in grey. We find that joint detection can
almost rule out all the wrong EoSs in the plot, while ET alone cannot. To be
more illuminative, we show 10\,$\sigma$ constraint at mass $m^{\rm S}_2 = 1.27
\, M_{\odot}$, which equals to the projected results of such a source  but at
$\sim$ 10 times further. We find that although many EoSs are within the
errorbar, multiband observations eliminate nearly half more of the EoSs than ET
alone. 

We have to admit that space-borne detector's help on reducing the errors on
$\Lambda$ is not comparable to a reduce in the luminosity distance or a change
of source location, as could be seen in Table~\ref{tab:result}. However, the
multiband improvement is a human endeavor other than Nature's choice about the
GW sources. With its help, we are one step closer to understand NS structure and
its EoS. \rep{In addition, we should be cautious when treating the uncertainties
on $\Lambda_{1,2}$, since our assumption is in the limit of $\delta\Lambda
\rightarrow 0$, which would break down when the mass ratio of two components
grows. In that case an estimation of $\delta\Lambda$ is needed, and if no prior
is provided such an estimation will lead to one-order-of-magnitude larger
uncertainty on $\Delta\Lambda_{1,2}$. But if some physically motivated prior
(e.g., from the universal relation and so on) is included, the effect will not
be large. Overall, our result should be viewed as reasonable but optimistic.
Meanwhile, as we have carefully checked, the multiband improvement factor
basically keeps the same level in both cases, with or without estimating
$\delta\Lambda$.}

%---------------------------------------------------------------------
\section{Summary}
\label{sec:sum}
%---------------------------------------------------------------------

In this paper, we adopted multiband detection strategy, namely using both
decihertz GW detectors and the ET, to jointly observe two classes of coalescing
binary systems. We take a GW170817-like BNS system and a GW200105-like NSBH
system as examples. We analyzed their PE uncertainties, presented the sky
distributions of the parameter constraints, and discussed the synergy effects in
detail. Here we give a brief summary on our key findings.
%--
\begin{enumerate}[(i)]
	\item Assuming the joint B-DECIGO and ET detection and adopting a PN
	waveform, we found that joint detection could break the strong correlation
	between the quadrupole parameter and spin parameter, which happens when the
	source is observed by a decihertz detector alone. The joint detection also
	breaks the strong correlations among localization parameters which happen
	when the source is observed by the ET alone. 
	\item We have showed that only the joint detection could effectively measure
	the quadrupole parameter for the BNS system. While tidal deformability is
	detectable by ET alone, joint detection still gives an improvement of a
	factor of 1--3. Tidal deformability measurements of NSBHs are dozens of
	times worse than that of BNSs, mainly because NSBHs' indistinctive tidal
	contribution to the GW phase. 
	\item Combining with the EoS information, we constrained the individual
	tidal deformability and demonstrated that multiband observations could rule
	out many more EoSs than the use of ET alone, which would greatly help to
	understand the behaviours of supranuclear dense matters at low temperature.
	\item We made comparisons of using different detectors and waveforms. We
	concluded that DECIGO, made up of 4 independent LISA-like detectors, is
	remarkably different from the other three space-borne decihertz detectors
	for its more complex design. We also showed that
	\texttt{IMRPhenomPv2\char`_NRTidalv2} waveform constrains the parameters
	looser than the PN waveform, but  the difference is small.
	\item BNS systems and some of the NSBH systems are expected to be
	accompanied with EM counterpart signals. Rapid alerts of source localization
	are crucial for successful multi-messenger follow-ups for some EM wavebands.
	We demonstrated that multiband detections could narrow down  the
	arcmin$^2$-level resolution from space-borne decihertz detectors alone by
	another dozens of times. Meanwhile, joint detections start to take effects
	about one day before the merger, which makes multiband pre-merger alerts
	even promising.
\end{enumerate}

In the first three observing runs of the LIGO/Virgo/KAGRA Collaborations, the
latencies of public alerts on GW candidates take minutes or hours after the preliminary
detection of GW events \citep{LIGOScientific:2019gag}. With the inclusion of
space-borne decihertz detectors, EM facilities can be fore-warned minutes or
hours before the merger with an accuracy in sky localization better than a
squared arcminute, which, will enable a deeper and more targeted search for the
EM counterpart by telescopes with narrow field of view, such as Swift-XRT
\citep{Burrows:2005gfa} and JWST \citep{Kalirai:2018qfg}. This hence will
maximize the science return in an unprecedented way.

Throughout the paper, priors are ignored in our calculations, or equivalently
priors are all considered uniform in each parameter, \rep{which leads to under-estimated parameter errors}. This can be improved if we
have more informed priors from astrophysical or other considerations, for
examples, from the limit of Kerr spin in general relativity or the universal relations for NSs. \rep{In addition, a full Bayesian analysis could incorporate various kind of prior knowledge and further test effects from different physical priors.}

\rep{The Fisher information matrix method, which can be viewed as a fast but approximate
version of parameter inference, however, has several caveats. It could be problematic when the parameter dimension grows \citep{Harry:2021hls}, 
in particular when including complex parameter corrections other than the
simple Gaussian ones \citep{Vallisneri:2007ev,Smith:2021bqc}. In that case, very
high correlations might appear and ruin the predictive ability of the Fisher
matrix. Though parameters in our parameter set are not highly correlated, the complex relation 
between intrinsic parameters, for example, between $\tilde\Lambda$ and $\delta\tilde\Lambda$, still makes us cautious viewing the results. 
Our analysis is thus only preliminary and indicative. Nevertheless, we have tested our method with the noise curve and system parameters of GW170817, and found consistent results for both tidal deformability and masses with what are reported in the dedicated LIGO/Virgo analysis~\citep{LIGOScientific:2017vwq, LIGOScientific:2018hze}. In addition, we have tested the validity of using the Fisher analysis in our scenarios using the likelihood ratio that was proposed in \citet{Vallisneri:2007ev}; we found that the linear signal approximation is satisfied, and the use of Fisher matrix is valid.} 
Moreover, we focus on the statistical 
errors of the multiband improvement. Nonetheless, such
enhancement could be hampered by the systematic errors which result from, e.g.
the inaccurate waveform modeling \citep{Isoyama:2018rjb,Gamba:2020wgg}.  A full
Bayesian analysis with consideration of systematic errors will help to address
these issues better.

In conclusion, multiband observations of BNS/NSBH systems will complement the
single waveband detections in providing new insights into nuclear matters under
extreme conditions, origins of high-energy astrophysical phenomena, and so on.
We foresee that with future multiband detection, our understanding of
astrophysics and fundamental physics will make  great progresses and pin down
unsolved important puzzles.

%---------------------------------------------------------------------
\begin{acknowledgements}
	We thank Yong Gao and Junjie Zhao for useful discussions, \rep{and the anonymous referee for critical comments}. This work was
	supported by the National Natural Science Foundation of China (11975027,
	11991053, 11721303),  the National SKA Program of China (2020SKA0120300),
	the Young Elite Scientists Sponsorship Program by the China Association for
	Science and Technology (2018QNRC001), the Max Planck Partner Group Program
	funded by the Max Planck Society, and the High-Performance Computing
	Platform of Peking University. Some of the results in this paper have been
	produced using the healpy and HEALPix packages \citep{2005ApJ...622..759G,
	Zonca2019}.
\end{acknowledgements}
%---------------------------------------------------------------------
\facilities{DECIGO, DO, ET}
\software{PyCBC \citep{Nitz:2020vym}, HEALPix \citep{2005ApJ...622..759G, Zonca2019}}

% \clearpage

%---------------------------------------------------------------------
\bibliography{nstidal}{}
\bibliographystyle{aasjournal}
%---------------------------------------------------------------------

\end{document}